\documentclass{jfm}

\usepackage{graphicx,color}
\usepackage{comment}
\usepackage{bm}
\usepackage{subcaption}
\usepackage{mathtools}
\usepackage{amsmath,amssymb}

\begin{document}

\newtheorem{lemma}{Lemma}
\newtheorem{corollary}{Corollary}

\shorttitle{Multi-scale steady solution for Rayleigh--B\'enard convection} %for header on odd pages
\shortauthor{S. Motoki, G. Kawahara and M. Shimizu} %for header on even pages

\title{Multi-scale steady solution for Rayleigh--B\'enard convection}
\author{Shingo Motoki\aff{1}\corresp{\email{motoki@me.es.osaka-u.ac.jp}}, Genta Kawahara\aff{1} \and Masaki Shimizu\aff{1}}
\affiliation{\aff{1}Graduate School of Engineering Science, Osaka University, 1-3 Machikaneyama, Toyonaka, Osaka 560-8531, Japan}

\maketitle

\begin{abstract}
We have found a multi-scale steady solution of the Boussinesq equations for Rayleigh--B\'enard convection in a three-dimensional periodic domain between horizontal plates with a constant temperature difference by using a homotopy from the wall-to-wall optimal transport solution given by Motoki {\it et al.} ({\it J. Fluid Mech.}, vol. 851, 2018, R4).
The connected steady solution, which turns out to be a consequence of bifurcation from a thermal conduction state at the Rayleigh number $Ra\sim 10^{3}$, is tracked up to $Ra\sim10^{7}$ by using a Newton--Krylov iteration.
The exact coherent thermal convection exhibits scaling $Nu\sim Ra^{0.31}$ (where $Nu$ is the Nusselt number) as well as multi-scale thermal plume and vortex structures, which are quite similar to those in the turbulent Rayleigh--B\'enard convection.
The mean temperature profiles and the root-mean-square of the temperature and velocity fluctuations are in good agreement with those of the turbulent states.
Furthermore, the energy spectrum follows Kolmogorov's $-5/3$ scaling law with a consistent prefactor, and the energy transfer to smaller scales in the wavenumber space agrees with the turbulent energy transfer.
\end{abstract}


\section{Introduction}\label{introduction}
Rayleigh--B\'enard convection, the buoyancy-driven flow in a horizontal fluid layer heated from below and cooled from above, is one of the most canonical flows widely observed in nature, including engineering materials.
The effect of buoyancy on a flow is characterised by the Rayleigh number $Ra$.
When $Ra$ exceeds a certain critical value $Ra_{c}$, the thermal conduction state becomes unstable, and two-dimensional (2D) steady convection rolls appear \citep{Drazin1981}.
At higher $Ra$, the convection becomes time-dependent, and subsequently exhibits turbulent states with multi-scale thermal and vortex structures.
One of the primary interests in the Rayleigh--B\'enard problem is the scaling of turbulent heat transfer with $Ra$, i.e., the dependence of the Nusselt number $Nu$ on $Ra$.
Over half a century ago, \cite{Malkus1954} derived the scaling $Nu\sim Ra^{1/3}$ by a marginal stability argument, based on the assumption that the thermal boundary layer adapts its thickness $\delta$ as $\delta/H\approx(Ra/Ra_{c})^{-1/3}$, where $H$ is the height of the fluid layer, so that the local $Ra$ in the boundary layer becomes marginally stable.
Subsequently, based on the mixing-length theory, \cite{Kraichnan1962} predicted a transition of the boundary layer from laminar to turbulent state, and derived the asymptotic scaling, $Nu\sim Ra^{1/2}$, with a logarithmic correction for very high $Ra$.
The scaling $Nu\sim Ra^{1/2}$ is known as ultimate scaling, and has been obtained as the rigorous upper bound on $Nu$ by variational approaches \citep{Doering1996,Plasting2003}.
In conventional turbulent Rayleigh--B\'enard convection, however, the ultimate scaling has not been observed yet.
A prominent experiment by \cite{Niemela2000} for very high $Ra$ exhibits $Nu\sim Ra^{0.31}$ even at $Ra\sim10^{17}$.
\cite{Grossmann2000} proposed a unifying scaling theory (GL theory) of global properties for $Ra$ and the Prandtl number $Pr$, based on decomposing the total scalar and energy dissipation into contributions from the bulk region and the boundary layer.
A lot of experiments and numerical simulations have demonstrated the validity of this theory \citep{Ahlers2009,Stevens2013}.
Per the theory, the scaling $Nu\sim Ra^{1/3}$ is derived in the high $Ra$ regime $10^{8}\lesssim Ra\lesssim10^{14}$ for $Pr\sim1$.
The transition to the ultimate scaling is also predicted for $Ra\gtrsim10^{14}$; however, the effective scaling is $Nu\sim Ra^{0.38}$ due to logarithmic corrections \citep{Grossmann2011}.
Although some results have shown the transition to $Nu\sim Ra^{0.38}$, the high-$Ra$ scaling is still being discussed \citep{Chilla2012,Zhu2018}.
On the other hand, for $10^{8}\lesssim Ra\lesssim10^{11}$, a lot of turbulent data exhibit $Nu\sim Ra^{0.31}$ \citep[see, e.g.][]{Niemela2006,He2012}.

\begin{figure}
\centering
	\begin{minipage}{.8\linewidth}
	\includegraphics[clip,width=\linewidth]{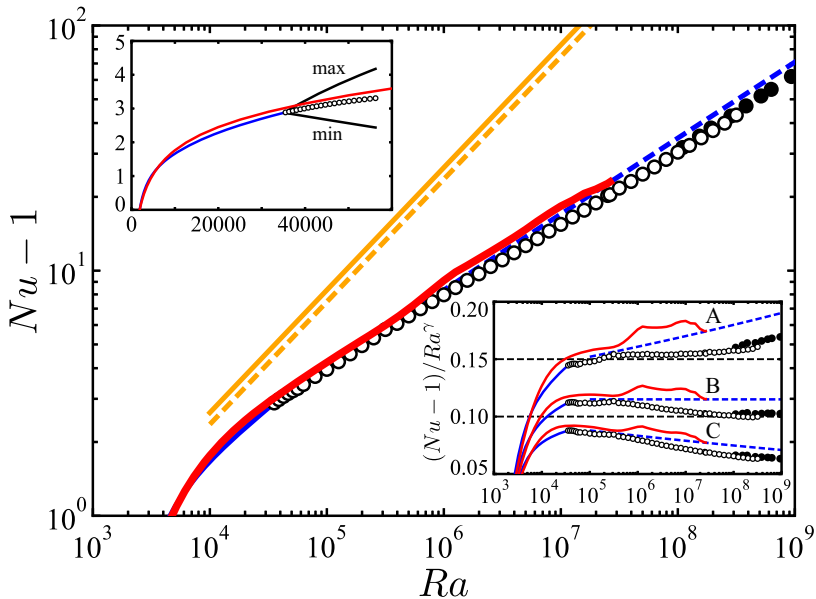}
	\end{minipage}
\caption{Nusselt number $Nu$ as a function of Rayleigh number $Ra$.
The red and blue solid lines represent the 3D and 2D steady solutions bifurcating from the conduction state at $Ra\approx1879$, respectively.
The open circles are the present turbulent data obtained in the horizontally square periodic domain, and the filled ones are the experimental turbulent data in a cylindrical container \citep{Niemela2006}.
The blue dashed line indicates the optimal scaling in the 2D steady solutions, $Nu-1=0.115Ra^{0.31}$ \citep{Waleffe2015,Sondak2015}.
The orange solid and dashed lines indicate the upper bound $Nu-1=0.02634Ra^{1/2}$ \citep{Plasting2003} and the optimal scaling $Nu-1=0.0236Ra^{1/2}$, respectively, evaluated from the wall-to-wall optimal transport states \citep{Motoki2018b}.
The black curves in the top-left inset show the maximal and minimum values of $Nu$ in the 3D time-periodic solution.
The bottom right inset shows $Nu$ compensated by $Ra^{\gamma}$: $\gamma=2/7$ (plot A); $\gamma=0.31$ (plot B); $\gamma=1/3$ (plot C). 
\label{fig:nu-ra}}
\end{figure}

Recently, Waleffe {\it et al.} \citep{Waleffe2015,Sondak2015} found a scaling $Nu=0.115Ra^{0.31}$, which is quite similar to the turbulent data fit $Nu=0.105Ra^{0.312}$ \citep{He2012} in 2D steady  Rayleigh--B\'enard convection for $10^7\lesssim Ra\lesssim10^9$.
They obtained optimal 2D steady solutions to maximise $Nu$ by changing the horizontal periods, and the scaling was achieved by a family of 2D solutions with the horizontal period that decreases with increasing $Ra$.
Although the result suggests that simple and coherent structures can capture the essence of turbulent convection, it does not imply that just any single 2D steady solution with a fixed horizontal period (maximal wavelength) can do it.
More recently, the wall-to-wall optimal transport problem, which is a variational problem of finding a divergence-free velocity field optimising scalar transport between two parallel plates, has been discussed \citep{Hassanzadeh2014,Tobasco2017}, and \cite{Motoki2018b} found three-dimensional (3D) steady velocity fields to be the optimal states maximising heat transfer between two isothermal no-slip parallel plates under the constraint of fixed total enstrophy.
The optimal states exhibit ultimate scaling, which is quite close to the rigorous upper bound $Nu-1=0.02634Ra^{1/2}$ \citep{Plasting2003}, as well as hierarchical self-similar vortex structures.
In 2D velocity fields, however, such multi-scale structures have not been observed to be the optimal state \citep{Souza2020}.
Although a 3D optimal state needs an external body force other than buoyancy, we proved that the optimal state can be continuously connected to a steady solution of the Boussinesq equation by using the homotopy continuation method (see appendix \ref{sec:homotopy}).
In the Rayleigh--B\'enard convection between horizontal boundaries, a 3D steady solution with convection cells also bifurcates from the conduction state at the same critical value of $Ra$ as the 2D steady solution (see the upper-left inset in Figure \ref{fig:nu-ra}), as convection rolls in any horizontal direction can exist simultaneously.
The connected solution is the 3D steady solution.
Although this solution is not stable, it exists even at high $Ra$.
In this paper, we demonstrate the ability of the invariant solution to capture key statistical features, as well as coherent thermal and flow structures in the turbulent Rayleigh--B\'enard convection. We then discuss the hierarchical multi-scale vortex structures and energy transfer.

The remainder of the paper is organised as follows.
In \S \ref{sec:system} we introduce the governing equations, the boundary conditions and the dimensionless parameters to characterise the thermal convection, and describe the numerical procedures to obtain nonlinear solutions.
The statistical properties and spatial structures of the 3D steady solution are presented in \S \ref{sec:3d}, and the hierarchical vortex structures are discussed in \S \ref{sec:hierarchy}.
Finally, summary and conclusions are presented in \S \ref{sec:summary}.
In the appendix \ref{sec:homotopy}, we present the homotopy continuation analysis from the optimal solution of the Euler--Lagrange equations for the wall-to-wall optimal transport problem, to the present 3D steady solution of the Boussinesq equations.
The parameter dependence of the 3D steady solution and the adequacy of the spatial resolution are shown in appendix \ref{sec:depend}.

\section{Boussinesq equations and numerical methods}\label{sec:system}
Let us consider a fluid layer between two horizontal plates heated from below and cooled from above, and employ the Oberbeck--Boussinesq approximation, wherein the density variations are only significant in the buoyancy term.
The time evolution of velocity field $\textit{\textbf{u}}(\textit{\textbf{x}},t)=u\textit{\textbf{e}}_{x}+v\textit{\textbf{e}}_{y}+w\textit{\textbf{e}}_{z}$ and temperature field $T(\textit{\textbf{x}},t)$ are described by the Boussinesq equations
\begin{eqnarray}
\label{eq:co}
\displaystyle
\nabla\cdot\textit{\textbf{u}}&=&0,\\
\label{eq:ns}
\displaystyle
\frac{\partial \textit{\textbf{u}}}{\partial t}+(\textit{\textbf{u}}\cdot\nabla)\textit{\textbf{u}}&=&-\frac{1}{\rho}\nabla p+\nu\nabla^{2}\textit{\textbf{u}}+g\alpha T\textit{\textbf{e}}_{z},\\
\label{eq:ad}
\displaystyle
\frac{\partial T}{\partial t}+(\textit{\textbf{u}}\cdot\nabla)T&=&\kappa\nabla^{2}T,
\end{eqnarray}
where $p(\textit{\textbf{x}},t)$ is pressure, and $\rho, \nu, g, \alpha$ and $\kappa$ are the mass density, kinematic viscosity, acceleration due to gravity, volumetric thermal expansivity, and thermal diffusivity, respectively.
$\textit{\textbf{e}}_{x}$ and $\textit{\textbf{e}}_{y}$ are mutually orthogonal unit vectors in the horizontal directions, while $\textit{\textbf{e}}_{z}$ is a unit vector in the vertical direction.
The two horizontal plates are positioned at $z=0$ and $z=H$, and the top (or bottom) wall surface is no-slip and impermeable, and held at a lower (or higher) constant temperature:
\begin{eqnarray}
\label{eq:bcs}
\displaystyle
\mbox{\boldmath$u$}(z=0)=\mbox{\boldmath$u$}(z=H)=\mbox{\boldmath$0$};\hspace{1em}T(z=0)=\Delta T>0,\hspace{1em}T(z=H)=0.
\end{eqnarray}
The velocity and temperature fields are supposed to be periodic in the $x$- and $y$-directions with the same period $L_{x}=L_{y}=L$.
The thermal convection is characterised by the Rayleigh number $Ra$ and the Prandtl number $Pr$,
\begin{eqnarray}
\label{eq:raprga}
\displaystyle
Ra=\frac{g\alpha\Delta T H^{3}}{\nu\kappa},\hspace{1em}Pr=\frac{\nu}{\kappa}.
\end{eqnarray}
The vertical heat flux is quantified by the Nusselt number
\begin{eqnarray}
\label{eq:nusselt}
\displaystyle
Nu=\frac{-\kappa{\langle \partial T/\partial z \rangle}_{xyt}+{\langle wT \rangle}_{xyt}}{\kappa\Delta T/H}=1+\frac{H}{\kappa\Delta T}{\left< wT \right>}_{xyzt},
\end{eqnarray}
where ${\langle \cdot \rangle}_{xyt}$ and  ${\langle \cdot \rangle}_{xyzt}$ represent the horizontal and time average and the volume and time average, respectively.
The second equality is given by the volume and time average of the equation (\ref{eq:ad}).

The equations (\ref{eq:co})--(\ref{eq:ad}) are discretised by employing a spectral Galerkin method based on the Fourier series expansion in the periodic horizontal directions and the Chebyshev polynomial expansion in the vertical direction.
The nonlinear terms are evaluated using a spectral collocation method.
The aliasing errors are removed with the aid of the $2/3$ rule and the $1/2$ rule for the Fourier transform and the Chebyshev transform, respectively.
Time advancement is performed with the Crank--Nicholson scheme and the second-order Adams--Bashforth scheme for the diffusion terms and the rest, respectively.
The nonlinear steady solutions are obtained by the Newton--Krylov iteration \citep[for more details, see \S 3 and appendix A in][]{Motoki2018a}.

In this paper, we present the steady solution and the turbulent states in the horizontally square periodic domain with $L/H=\pi/2\approx1.57$ for $Pr=1$.
The domain is the same as that of the optimal states derived by \cite{Motoki2018b}.
The numerical process is carried out on $128^{3}$ grid points for $Ra<10^7$ and $256^{3}$ grid points for $Ra\ge10^7$.
In the domains with $L/H=2\pi/3.117\approx2.02$ and $1$, and for $Pr=7$, we confirm that the effects of the domain size, $Pr$ and the spatial resolution on the heat flux at high $Ra$ as well as the thermal and flow structures in 3D steady solutions, which will be discussed in the following sections, are insignificant in appendix \ref{sec:depend}.

All the 3D steady solutions presented in this paper satisfy the $\pi/2$-rotation symmetry
\begin{eqnarray}
\label{eq:sym_rot}
[u,v,w,T](x,y,z)&=&[v,-u,w,T](y,-x,z),
\end{eqnarray}
as well as the mirror symmetry
\begin{eqnarray}
\label{eq:sym_mirror}
[u,v,w,T](x,y,z)&=&[-u,v,w,T](-x,y,z)\nonumber \\
&=&[u,-v,w,T](x,-y,z),
\end{eqnarray}
the shift-and-reflection symmetry
\begin{eqnarray}
\label{eq:sym_shiftref}
[u,v,w,T](x,y,z)&=&[u,v,-w,1-T](x+L/2,y+L/2,1-z).
\end{eqnarray}
Although these symmetries are not imposed explicitly, they are satisfied in the solutions at $Ra\lesssim10^{6}$.
At $Ra\sim10^{7}$ we directly impose the symmetries (\ref{eq:sym_rot})--(\ref{eq:sym_shiftref}) in order to reduce the computational degrees of freedom.

\begin{figure}
\centering
	\begin{minipage}{.38\linewidth}
	(\textit{a})\\
	\includegraphics[clip,width=\linewidth]{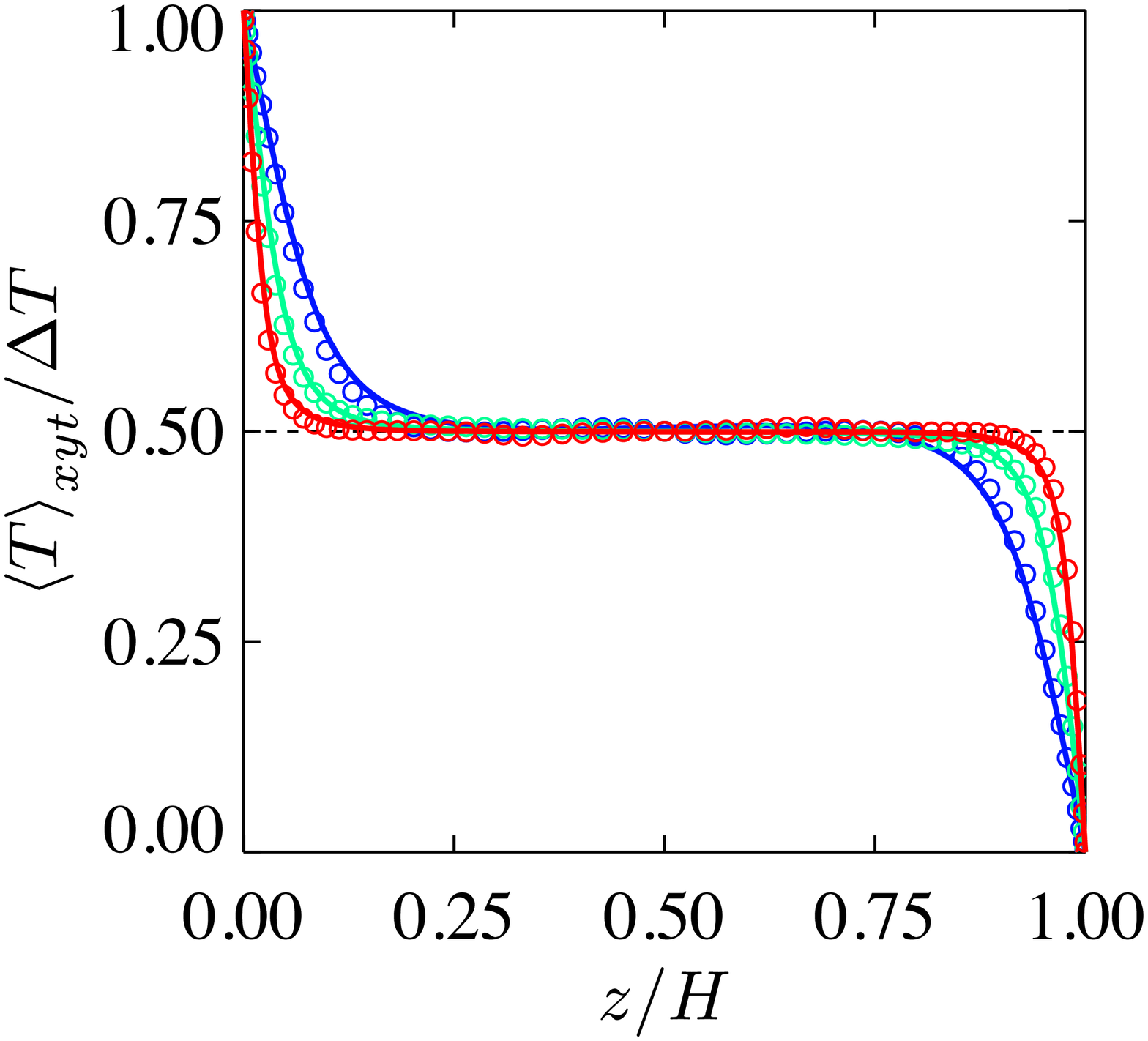}
	\end{minipage}
	\hspace{1em}
	\begin{minipage}{.38\linewidth}
	(\textit{b})\\
	\includegraphics[clip,width=\linewidth]{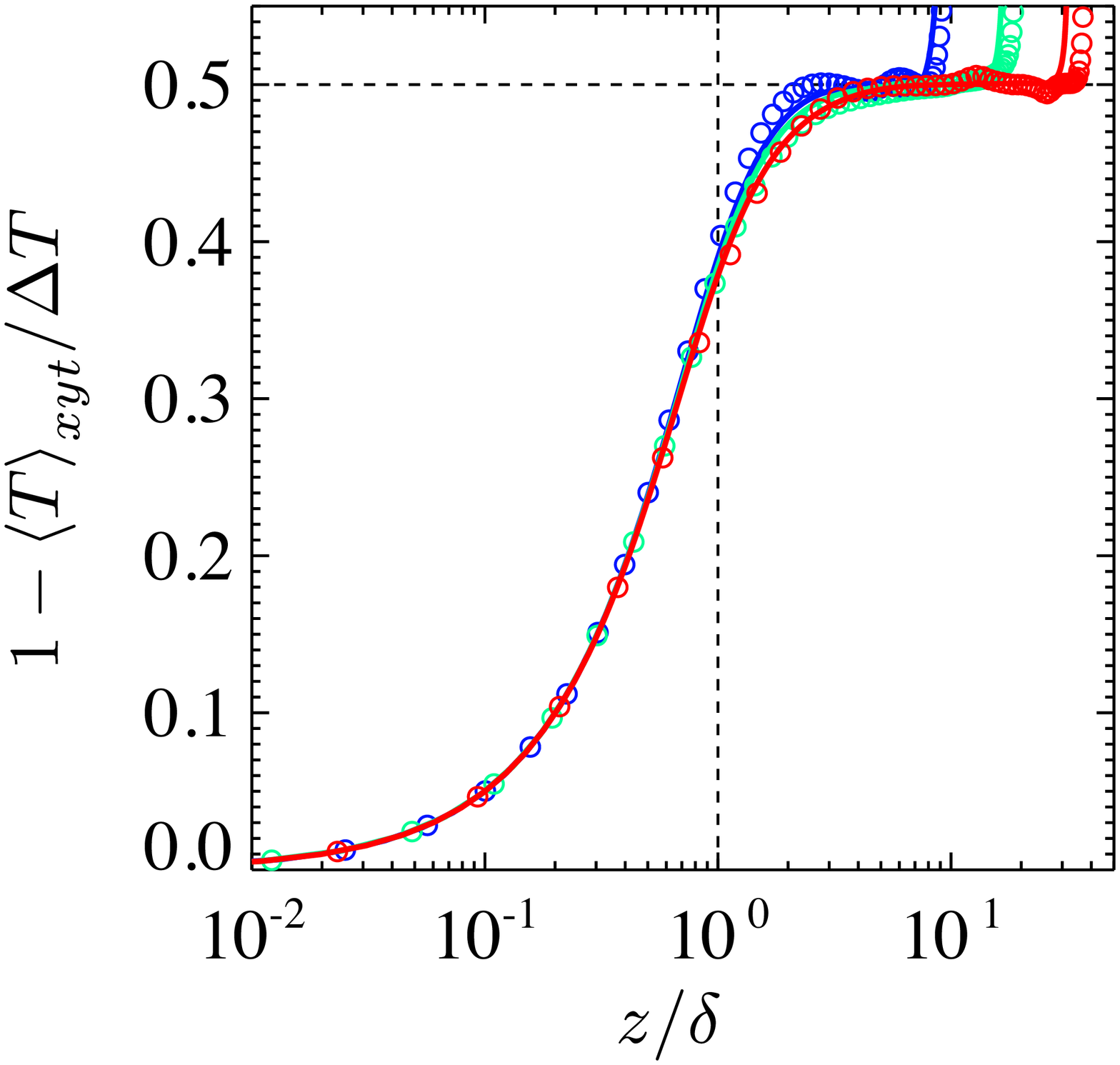}
        \end{minipage}

	\begin{minipage}{.38\linewidth}
	(\textit{c})\\
	\includegraphics[clip,width=\linewidth]{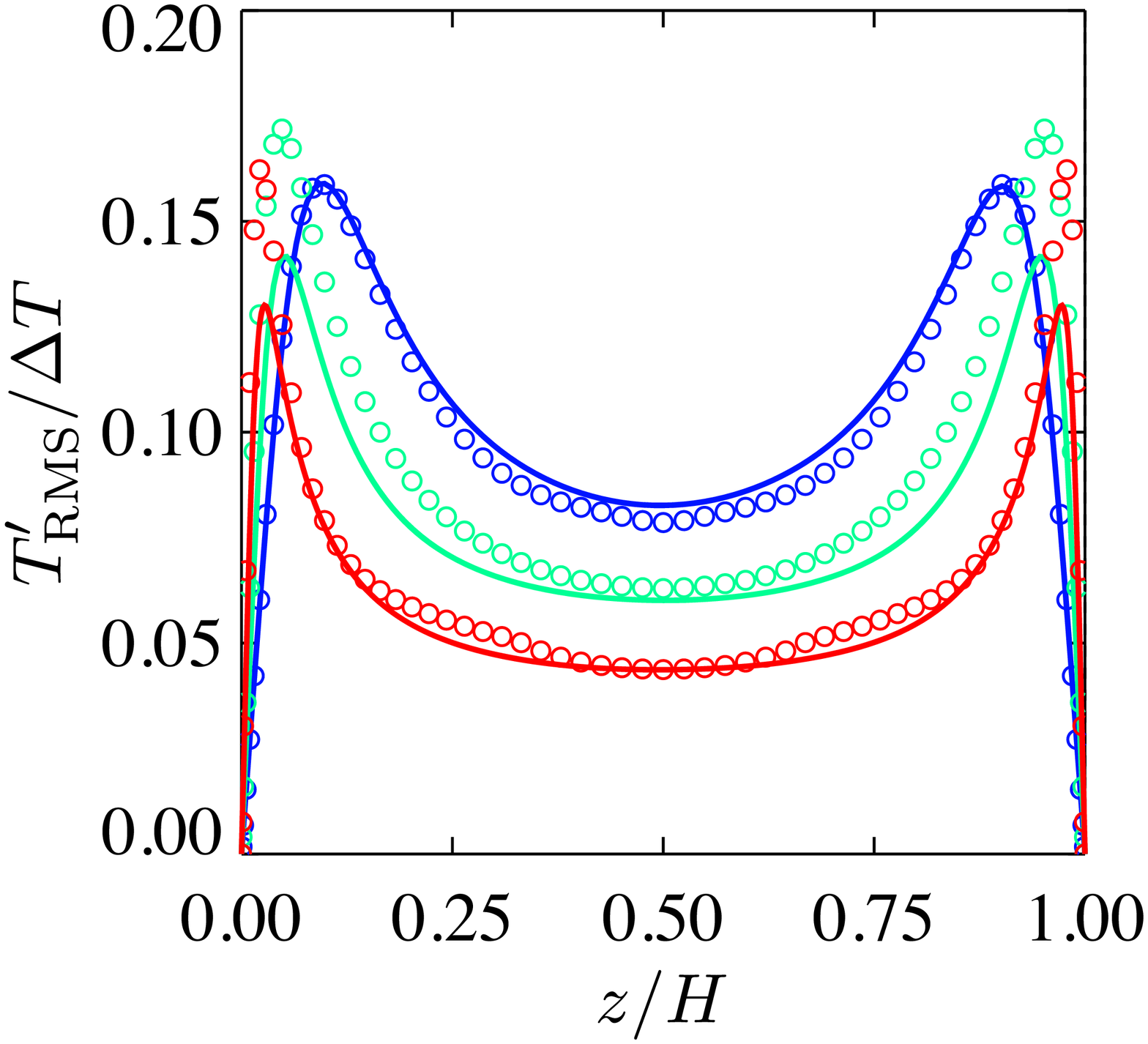}
	\end{minipage}
	\hspace{1em}
	\begin{minipage}{.38\linewidth}
	(\textit{d})\\
	\includegraphics[clip,width=\linewidth]{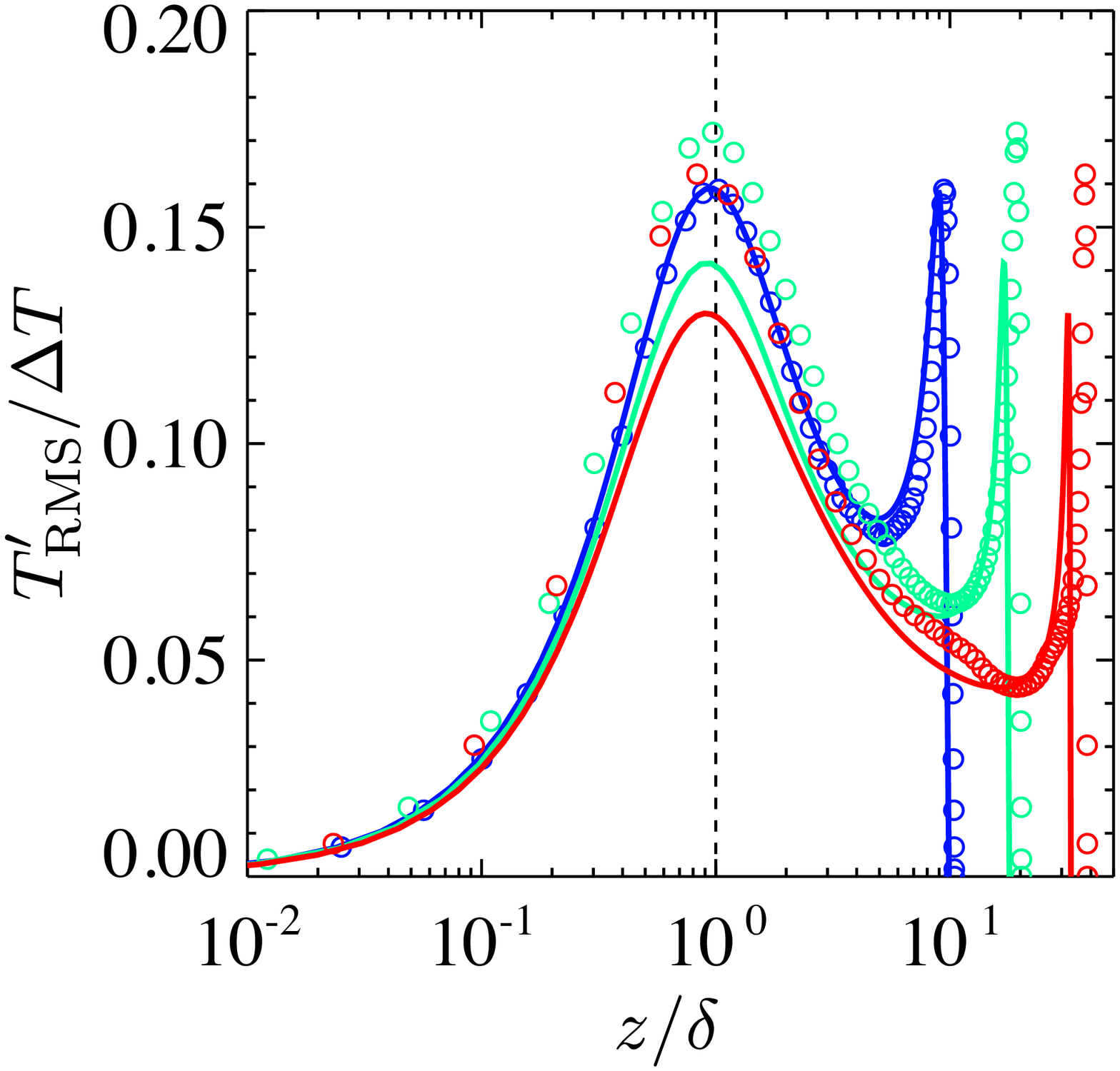}
	\end{minipage}

	\begin{minipage}{.38\linewidth}
	(\textit{e})\\
	\includegraphics[clip,width=\linewidth]{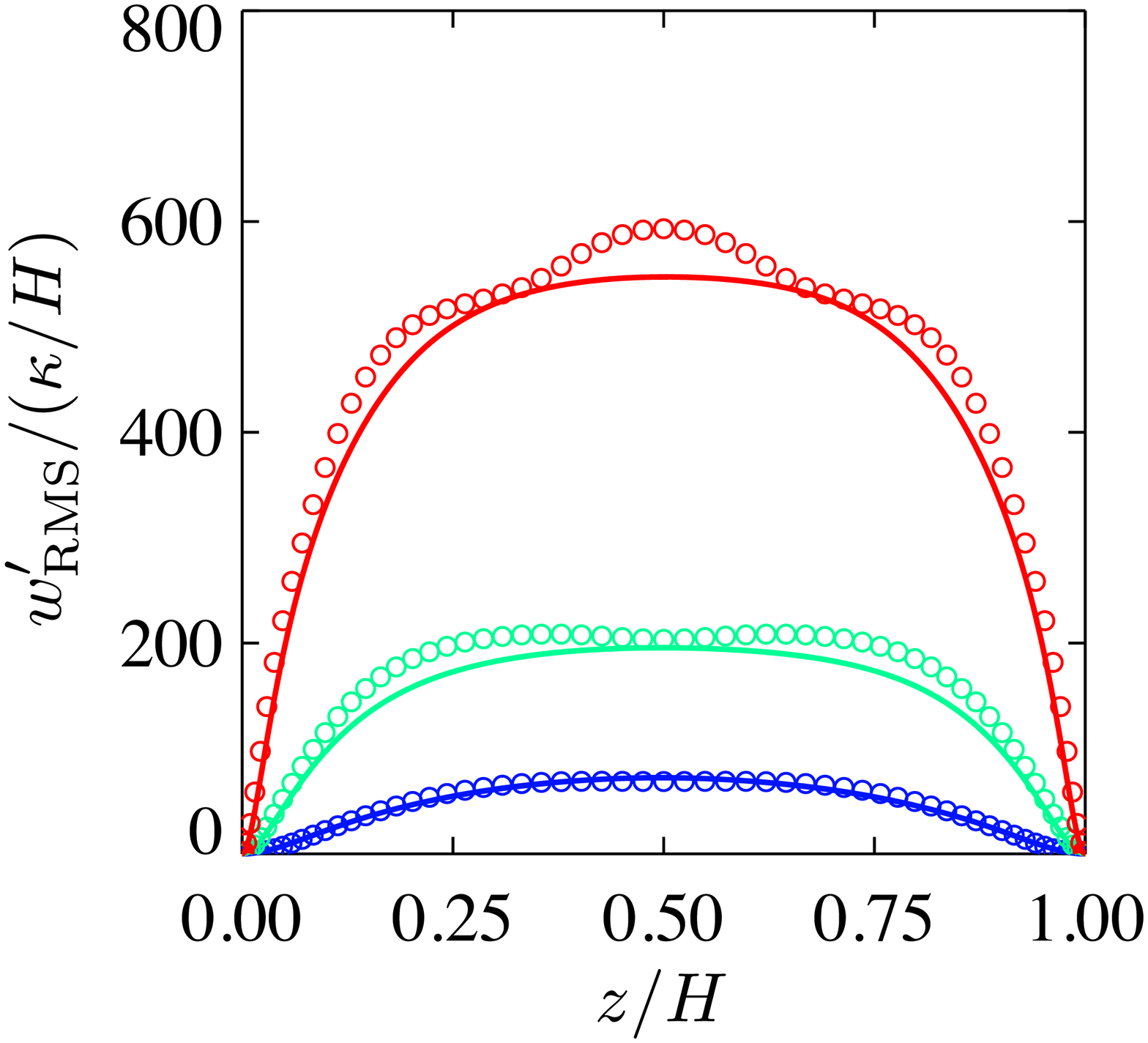}
	\end{minipage}
	\hspace{1em}
	\begin{minipage}{.38\linewidth}
	(\textit{f})\\
	\includegraphics[clip,width=\linewidth]{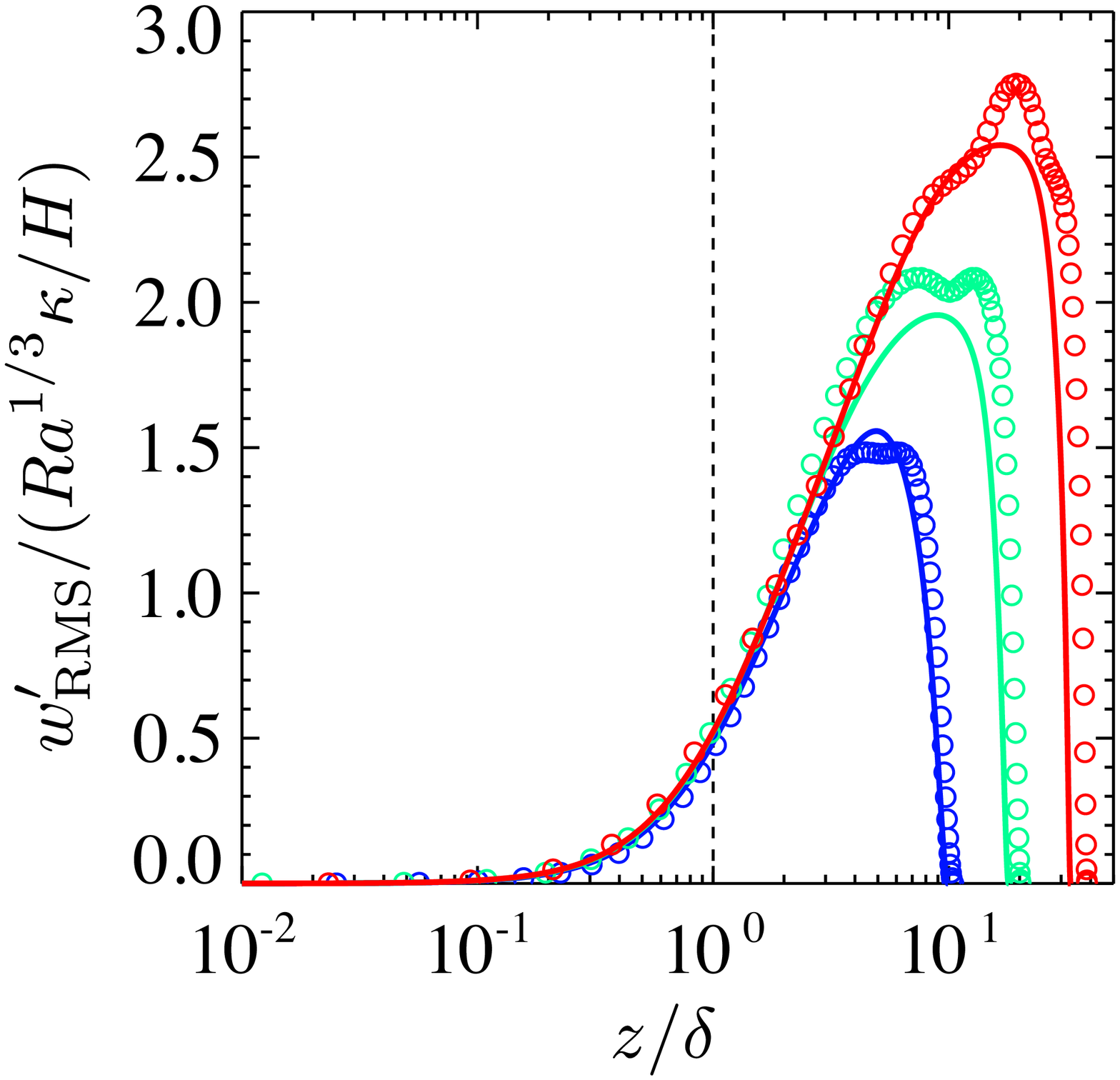}
	\end{minipage}
\caption{Mean temperature and RMS of the temperature and vertical velocity fluctuations as a function of (\textit{a,c,e}) $z/H$ and (\textit{b,d,f}) $z/\delta$ in the three-dimensional steady solution (circles) and the turbulent state (lines).
The blue, green and red plots are obtained at $Ra=10^5$, $Ra=10^6$ and $Ra=10^7$, respectively.
$\delta$ is the thermal conduction layer thickness scales as $\delta/H=1/(2Nu)$.
\label{fig:mte_rms}}
\end{figure}

\section{Three-dimensional steady solution}\label{sec:3d}
\subsection{$Nu$-$Ra$ scaling}\label{sec:scaling}
Figure \ref{fig:nu-ra} presents $Nu$ as a function of $Ra$.
The red line shows the 3D steady solution, and the open and filled circles represent the present turbulent data in the horizontally square periodic domain and the experimental data in a cylindrical container \citep{Niemela2006}, respectively.
The 3D steady solution maintains slightly larger $Nu$ than the turbulent states even at $Ra\sim10^{7}$.
The bottom right inset shows $Nu$ compensated by $Ra^{\gamma}$, and the scaling exponent $\gamma$ of turbulent states shows $2/7$ for $Ra\lesssim10^{7}$, and at higher $Ra$ it changes to $0.31$.
Such a transition has been experimentally and numerically observed for $Pr\sim1$ \citep{Castaing1989,Silano2010}.
Meanwhile, the exponent of heat flux in the 3D steady solution is greater than $2/7$ but less than $1/3$, and it can be approximated to $Nu-1=0.115Ra^{0.31}$ \citep[blue dashed,][]{Waleffe2015,Sondak2015}, which is achieved by a family of 2D steady solutions with optimal horizontal periods.
The orange dashed line indicates the optimal scaling $Nu-1=0.0236Ra^{1/2}$ \citep{Motoki2018b} given by the 3D optimal states in the wall-to-wall optimal transport problem, and it is quite close to the rigorous upper bound \citep{Plasting2003}.
Although optimal states exhibiting significantly high heat flux have been achieved by external body force being different from buoyancy, the steady solution can be continuously connected to the present 3D steady solution of the full Boussinesq equations by a homotopy from the body force to the buoyancy, as shown in appendix \ref{sec:homotopy}.

\subsection{Mean temperature and root-mean-square profiles}\label{sec:profiles}
The 3D steady solution reproduces the mean temperature of turbulent states in the whole region. Furthermore, the root-mean-square (RMS) values are also in good agreement with each other (figure \ref{fig:mte_rms}).
Note that the RMS values are obtained from the horizontal average for the steady solutions, and the time and horizontal averages for the turbulent states.
In the bulk region, all mean temperature profiles are flattened, as a result of the nearly complete mixing by large-scale convection.
The temperature difference $\Delta T/2$ exists only at the thermal conduction layer, $0\le z\lesssim2\delta/H=1/Nu$, and $T'_{\rm RMS}$ has peaks at $z/\delta\approx1$.
If the advection, diffusion, and buoyancy terms in the Navier--Stokes equation (\ref{eq:ns}) at the conduction layer are balanced as
\begin{eqnarray}
\displaystyle
\frac{w'^{2}}{\delta}\sim\nu\frac{w'}{\delta^{2}}\sim g\alpha\Delta T,
\end{eqnarray}
(the balance between the advection and diffusion terms is given by that in the energy equation (\ref{eq:ad}) for $\nu\sim\kappa$) then the near-wall vertical velocity would be
\begin{eqnarray}
\displaystyle
w'\sim\sqrt{g\alpha\Delta T\delta}\sim Ra^{1/3}\frac{\kappa}{H},
\end{eqnarray}
and yields the scaling law
\begin{eqnarray}
\displaystyle
Nu\sim Ra^{1/3},
\end{eqnarray}
which has been given by Malkus' theory \citep{Malkus1954} and the GL theory \citep{Grossmann2000}.
As shown in figure \ref{fig:mte_rms}(\textit{f}) the RMS vertical velocity $w'_{\rm RMS}$ scales as $Ra^{1/3}\kappa/H$ near the wall $z/\delta\sim1$.

\begin{figure}
\centering
	\begin{minipage}{.42\linewidth}
	(\textit{a})\\
	\includegraphics[clip,width=\linewidth]{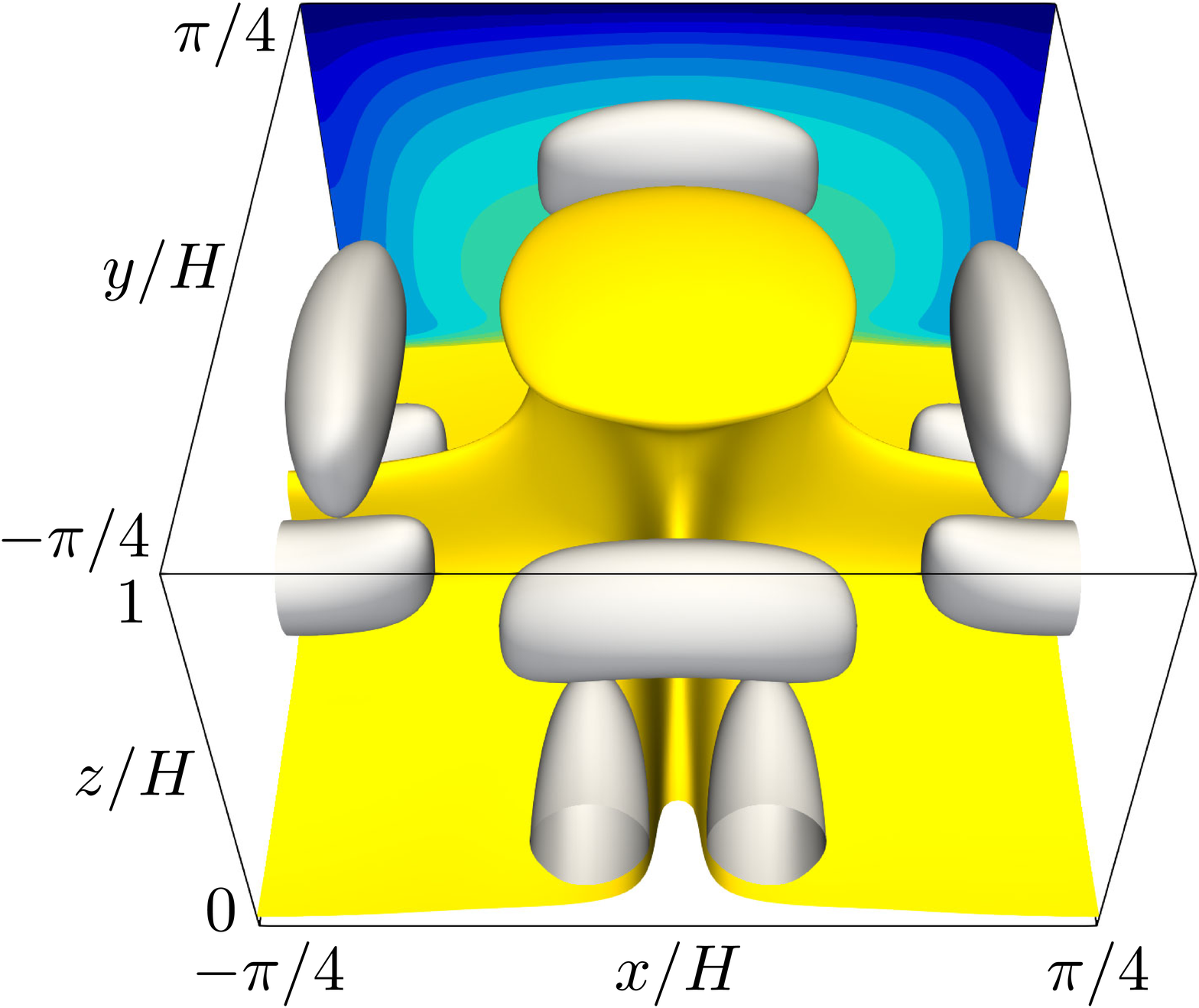}
	\end{minipage}
	\begin{minipage}{.42\linewidth}
	(\textit{b})\\
	\includegraphics[clip,width=\linewidth]{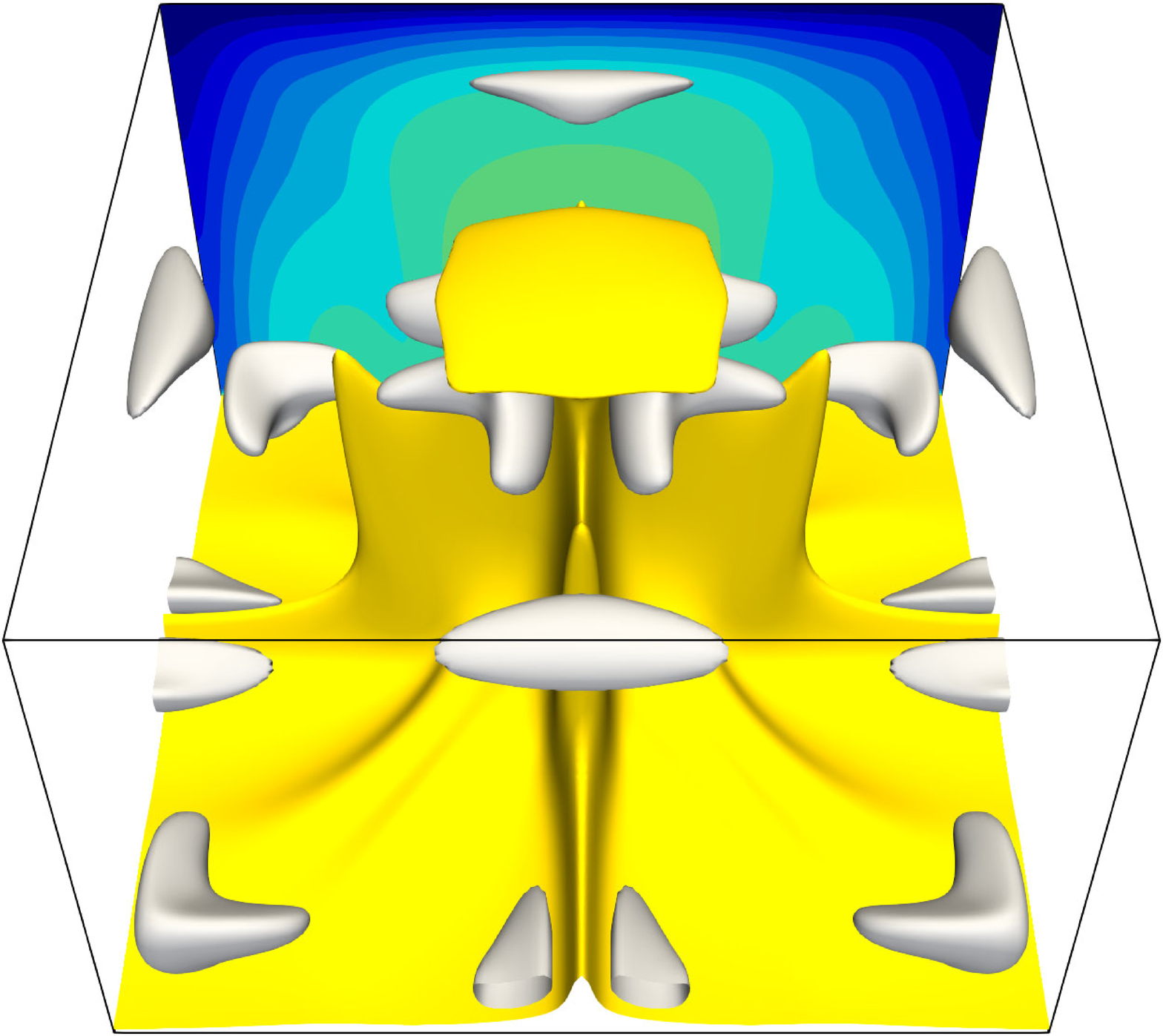}
        \end{minipage}

	\begin{minipage}{.42\linewidth}
	(\textit{c})\\
	\includegraphics[clip,width=\linewidth]{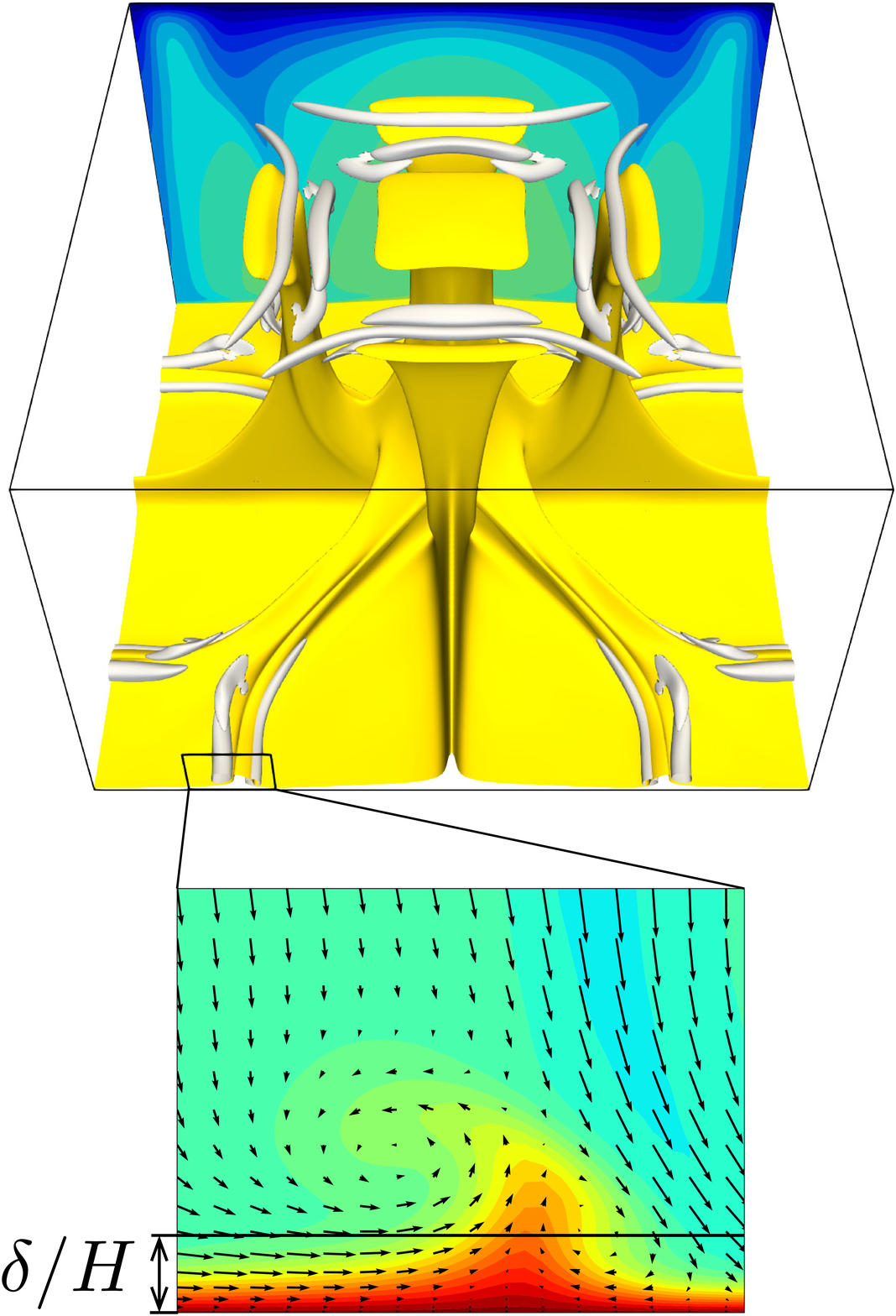}
	\end{minipage}
	\begin{minipage}{.42\linewidth}
	(\textit{d})\\
	\includegraphics[clip,width=\linewidth]{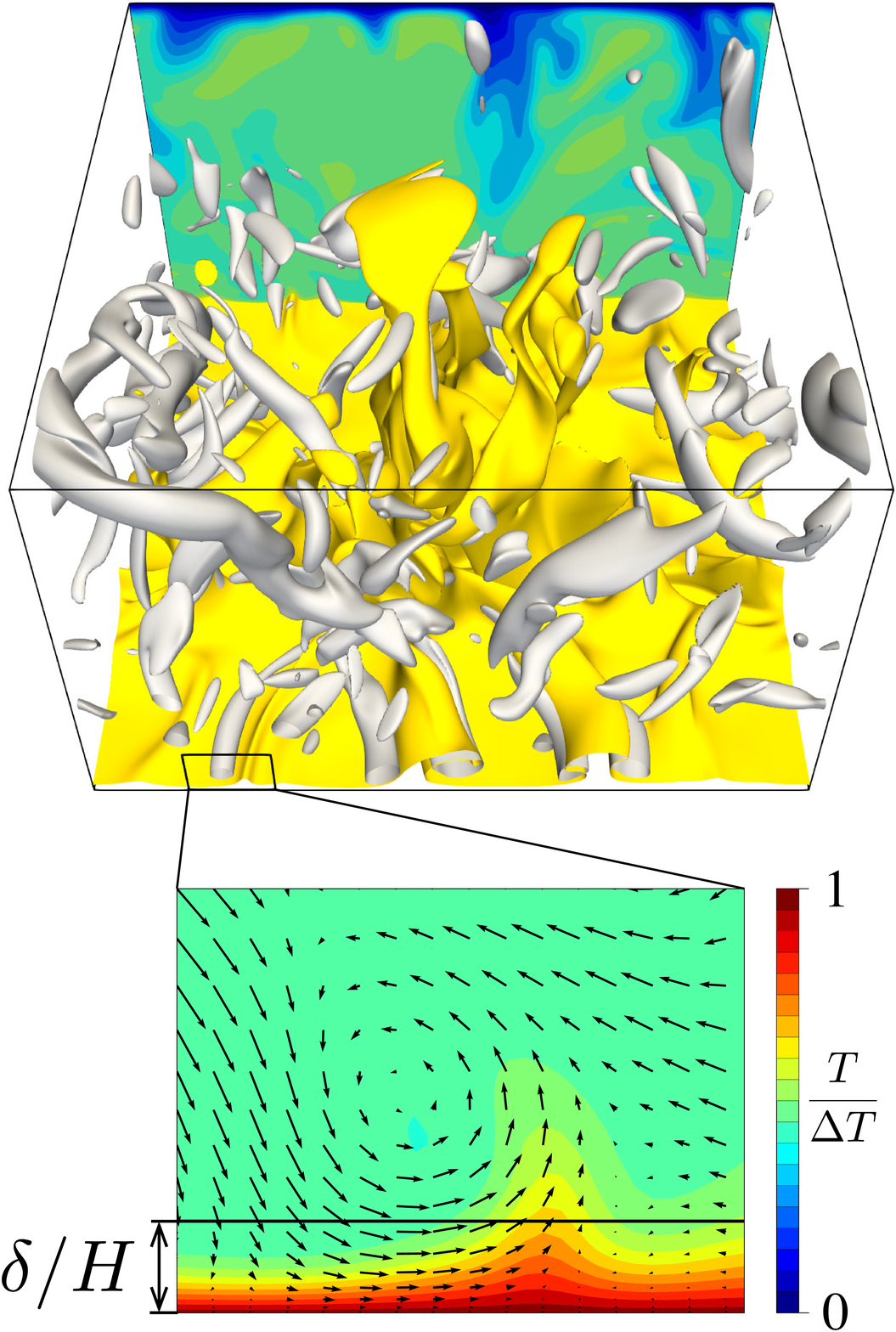}
	\end{minipage}
\caption{Thermal and flow structures in the 3D steady solution at (\textit{a}) $Ra=10^5$, (\textit{b}) $Ra=10^6$ and (\textit{c}) $Ra=10^7$, and (\textit{d}) the turbulent state at $Ra=10^7$.
The yellow and grey objects show the isosurfaces of the temperature $T/\Delta T=0.6$ and the positive second invariant of the velocity gradient tensor, (\textit{a}) $Q/(\kappa^2/H^4)=1.28\times10^5$, (\textit{b}) $Q/(\kappa^2/H^4)=2.4\times10^6$ and (\textit{c,d}) $Q/(\kappa^2/H^4)=8\times10^7$, respectively.
The contours represent temperature $T$ on the plane $y/H=\pi/4 (=-\pi/4)$, and the velocity vectors $(u,w)$ in the enlarged views in (\textit{c,d}) are superposed.
\label{fig:structure}}
\end{figure}

\subsection{Thermal and flow structures}\label{sec:structure}
Figure \ref{fig:structure} visualises the thermal and flow structures in the 3D steady solution and the turbulent state.
The yellow objects show the isosurfaces of temperature $T/\Delta T=0.6$, representing high-temperature plumes, and the grey objects display the vortex structures visualised by the positive second invariant of the velocity gradient tensor
\begin{eqnarray}
\displaystyle
Q=-\frac{1}{2}\frac{\partial u_{i}}{\partial x_{j}}\frac{\partial u_{j}}{\partial x_{i}}.
\end{eqnarray}
As $Ra$ in the 3D steady solution increases, smaller thermal plume structures (and relevant smaller and stronger tube-like vortex structures) appear near the walls without affecting the already existing large-scale structures.
At $Ra=10^7$ (figure \ref{fig:structure}\textit{c}), we observed sheet-like thermal plumes with smallest-scale vortices, which are quite similar with those observed in the snapshot of the turbulent state (figure \ref{fig:structure}\textit{d}).
The smallest-scale structures are generated in the thermal conduction layer, and the size of the plumes and vortices scale with the thickness $\delta$.
In 2D steady solutions \citep{Waleffe2015,Sondak2015}, the appearance of such small-scale plume and vortex structures has not been observed for a fixed horizontal period, and the scaling $Nu\sim Ra^{0.31}$ is achieved by a family of solutions with smaller horizontal periods as the $Ra$ increases.
It should be stressed that the single 3D steady solution spontaneously reproduces the multi-scale coherent structures of convective turbulence.

\section{Hierarchical vortices and energy transfer in wavenumber space }\label{sec:hierarchy}
\begin{figure}
\centering
	\begin{minipage}{.35\linewidth}
	(\textit{a})\\
	\includegraphics[clip,width=\linewidth]{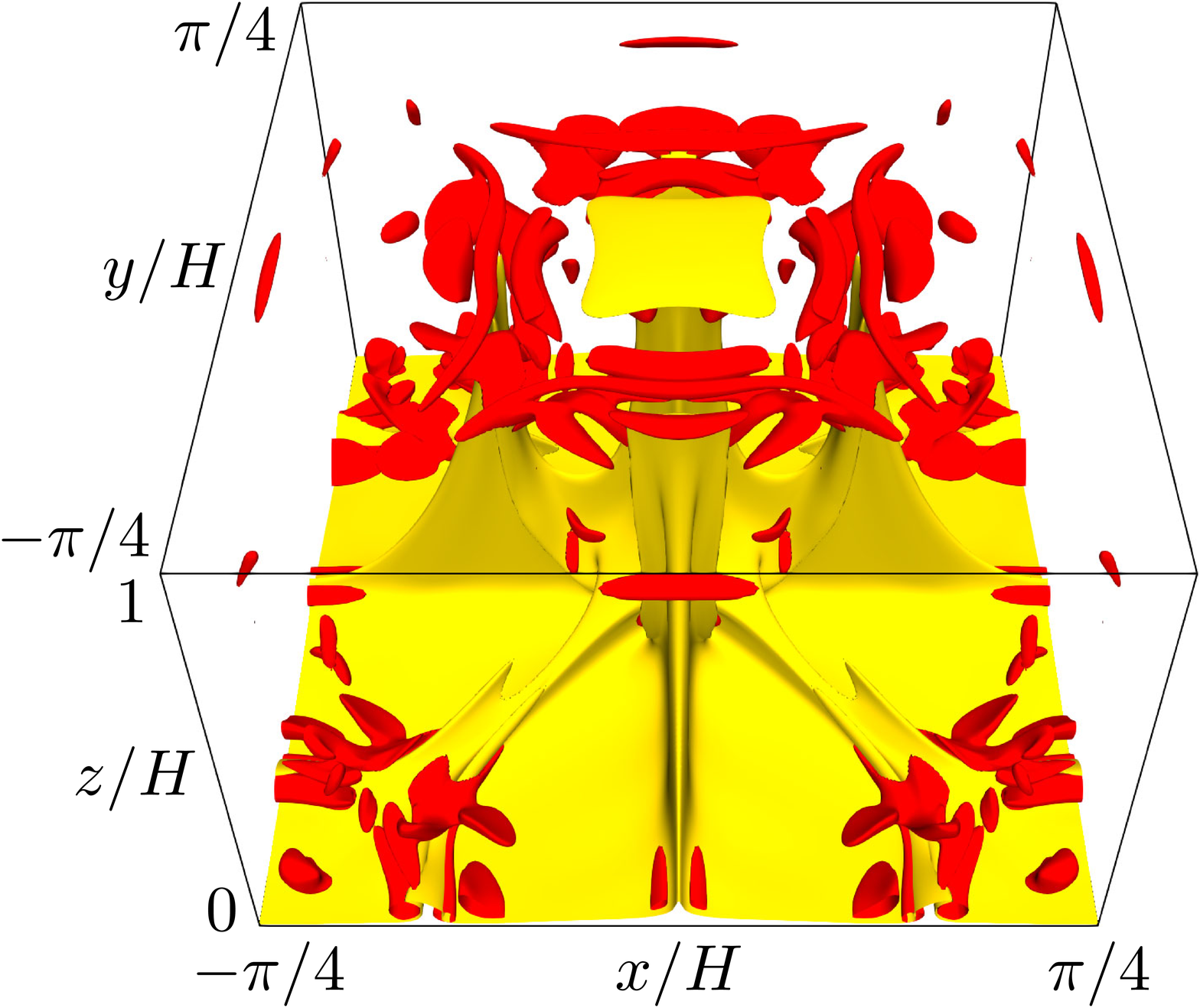}
	\end{minipage}
	\begin{minipage}{.35\linewidth}
	(\textit{b})\\
	\includegraphics[clip,width=\linewidth]{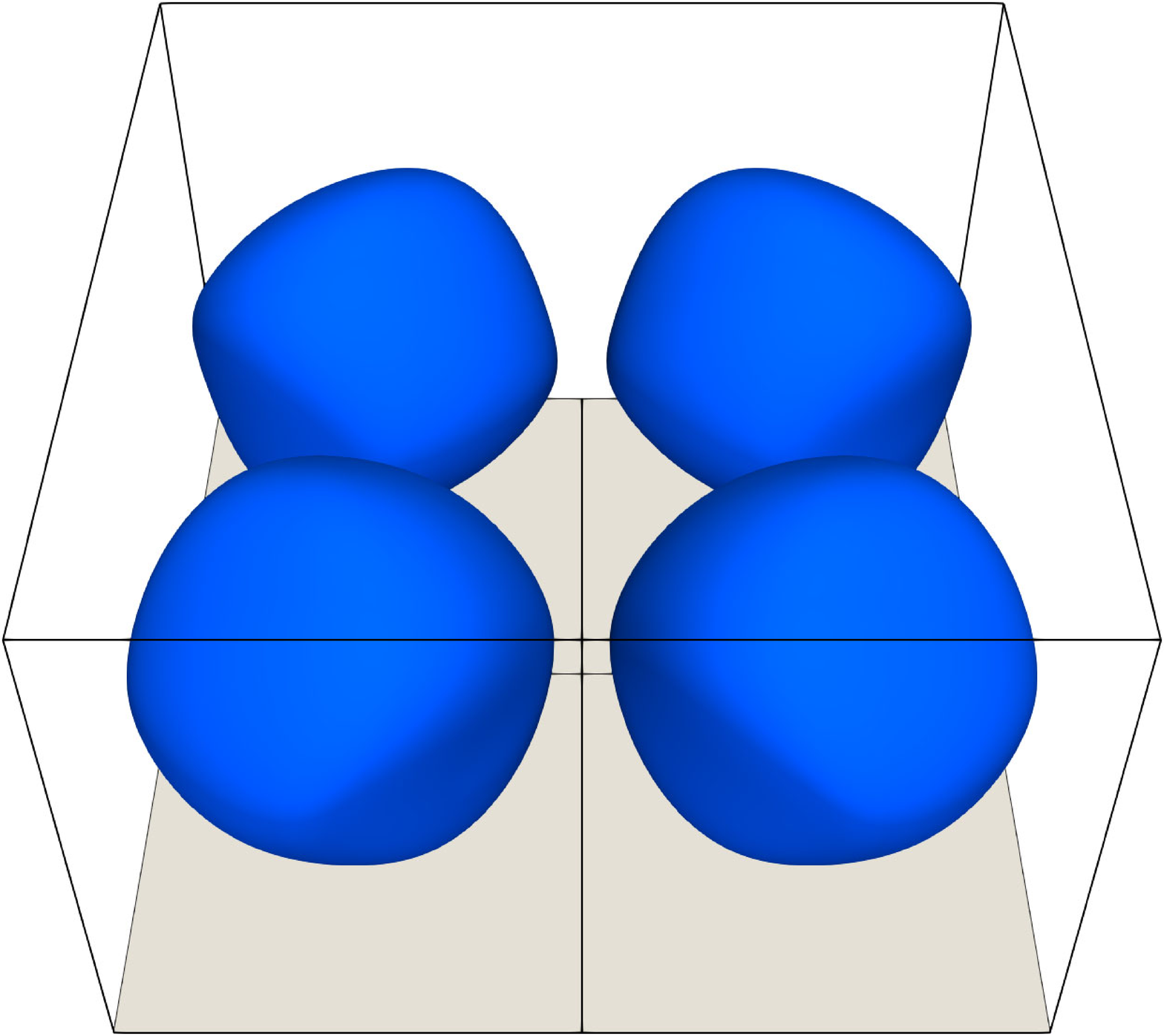}
        \end{minipage}

	\begin{minipage}{.35\linewidth}
	(\textit{c})\\
	\includegraphics[clip,width=\linewidth]{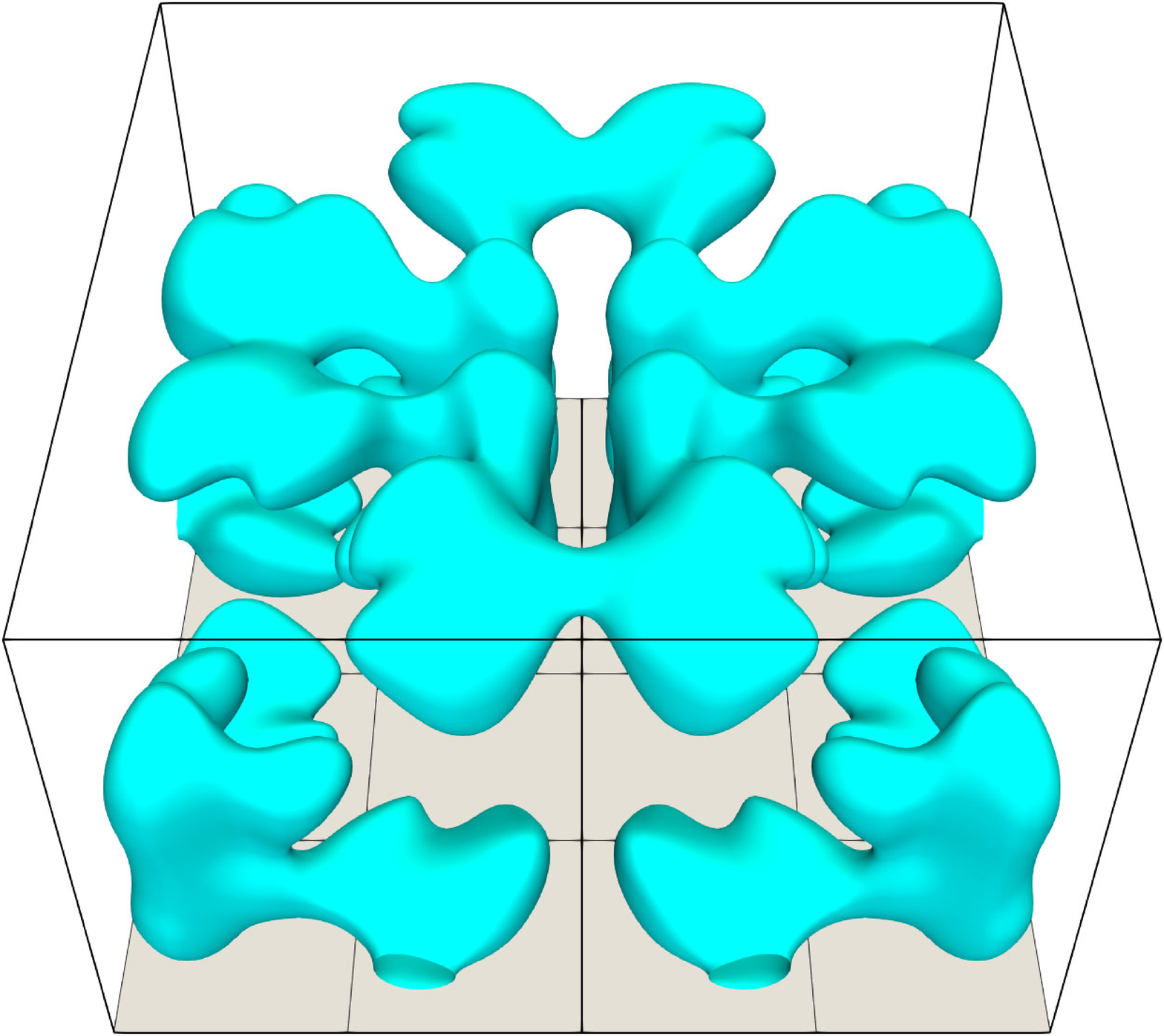}
	\end{minipage}
	\begin{minipage}{.35\linewidth}
	(\textit{d})\\
	\includegraphics[clip,width=\linewidth]{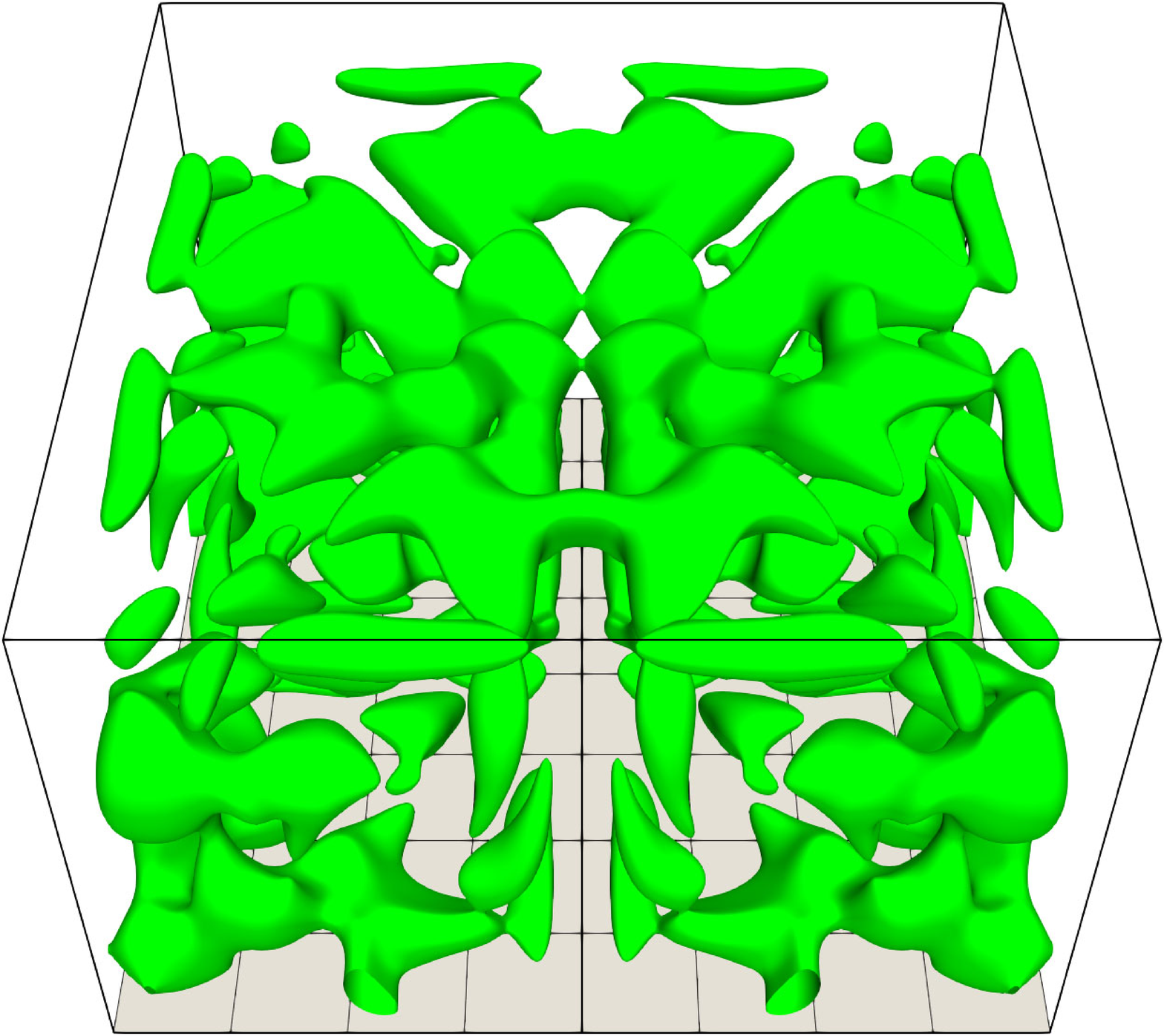}
	\end{minipage}
	
	\begin{minipage}{.35\linewidth}
	(\textit{e})\\
	\includegraphics[clip,width=\linewidth]{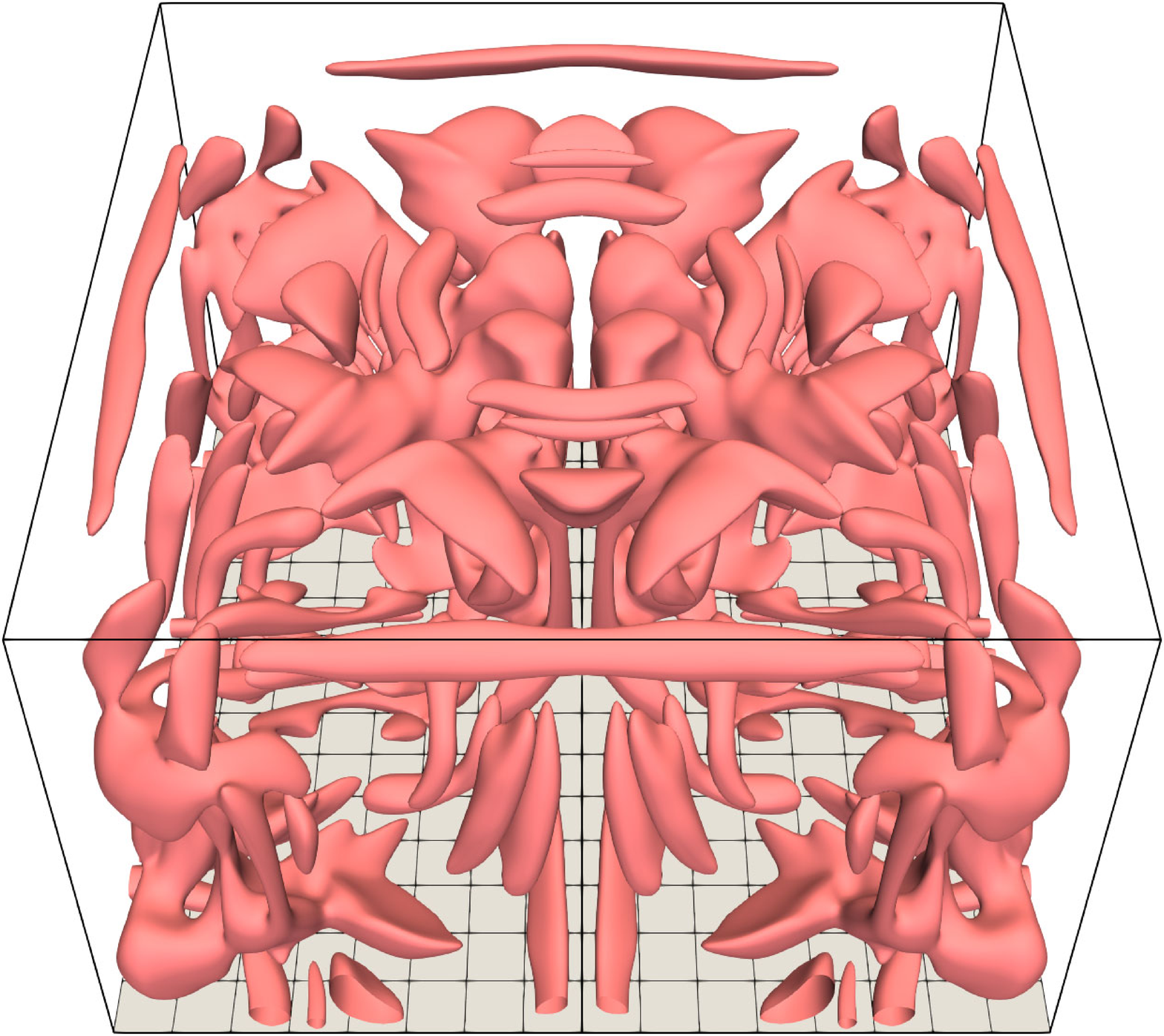}
	\end{minipage}
	\begin{minipage}{.35\linewidth}
	(\textit{f})\\
	\includegraphics[clip,width=\linewidth]{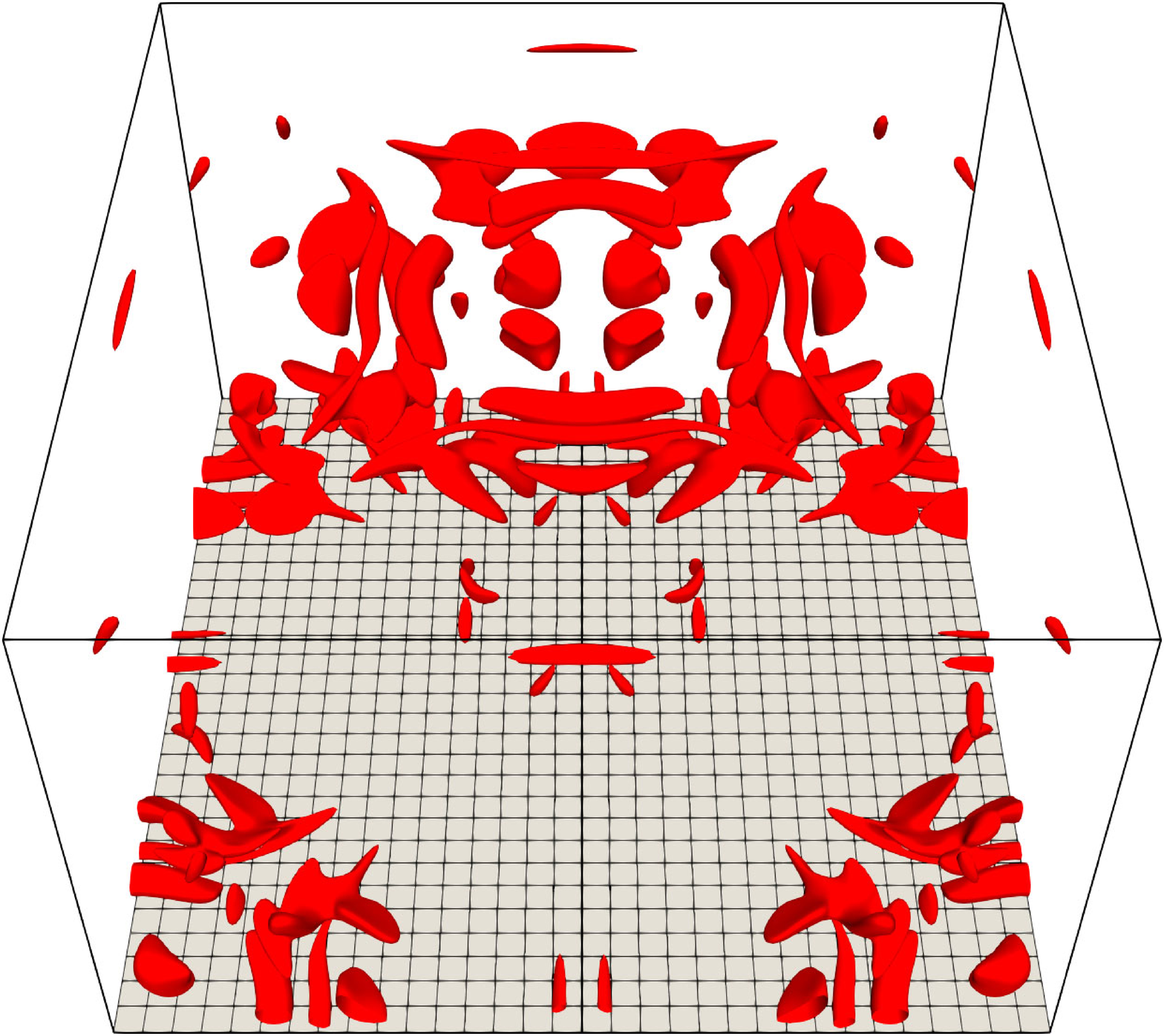}
	\end{minipage}
	
	\begin{minipage}{.35\linewidth}
	(\textit{g})\\
	\includegraphics[clip,width=\linewidth]{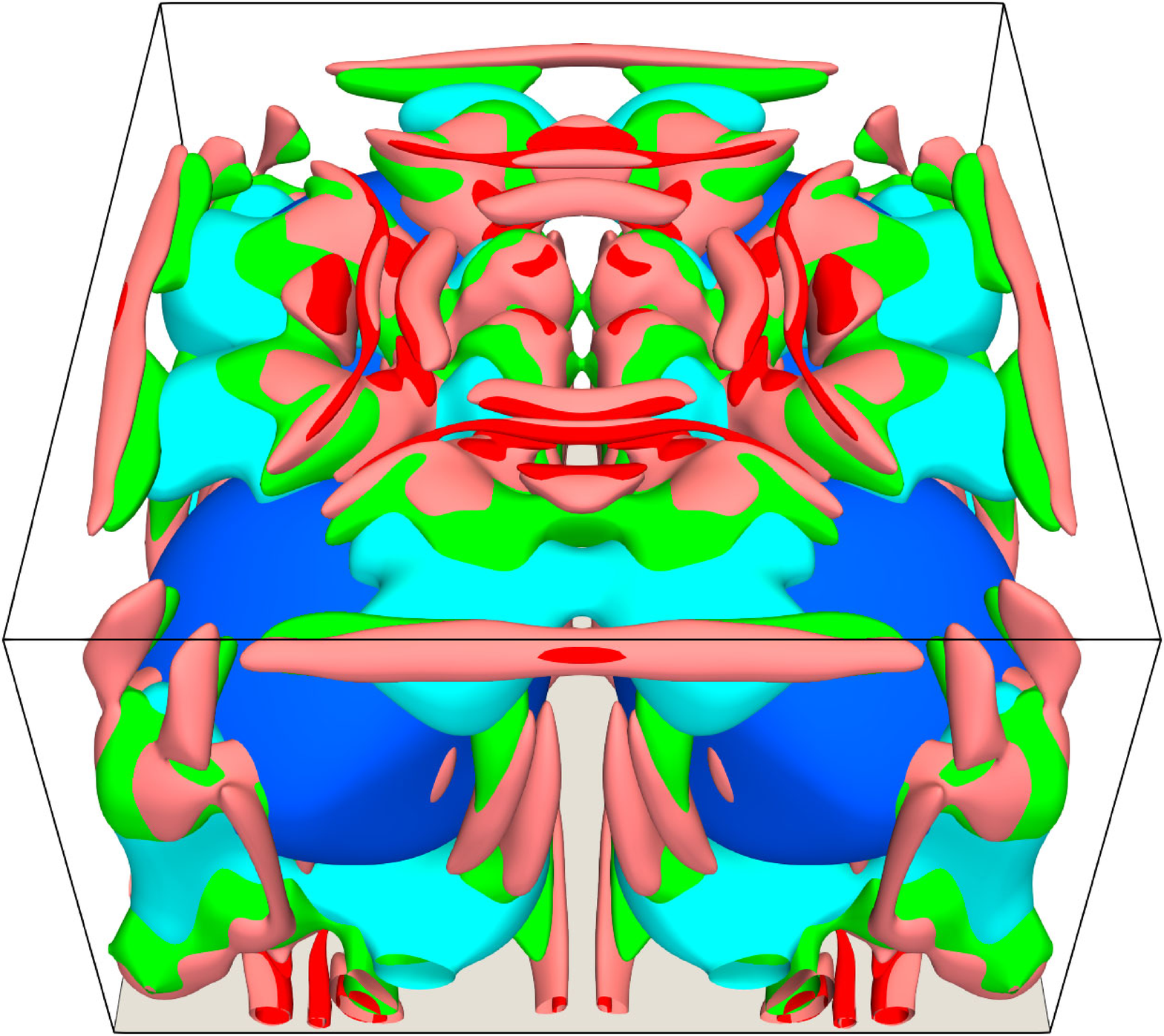}
	\end{minipage}
	\begin{minipage}{.35\linewidth}
	(\textit{h})\\
	\includegraphics[clip,width=\linewidth]{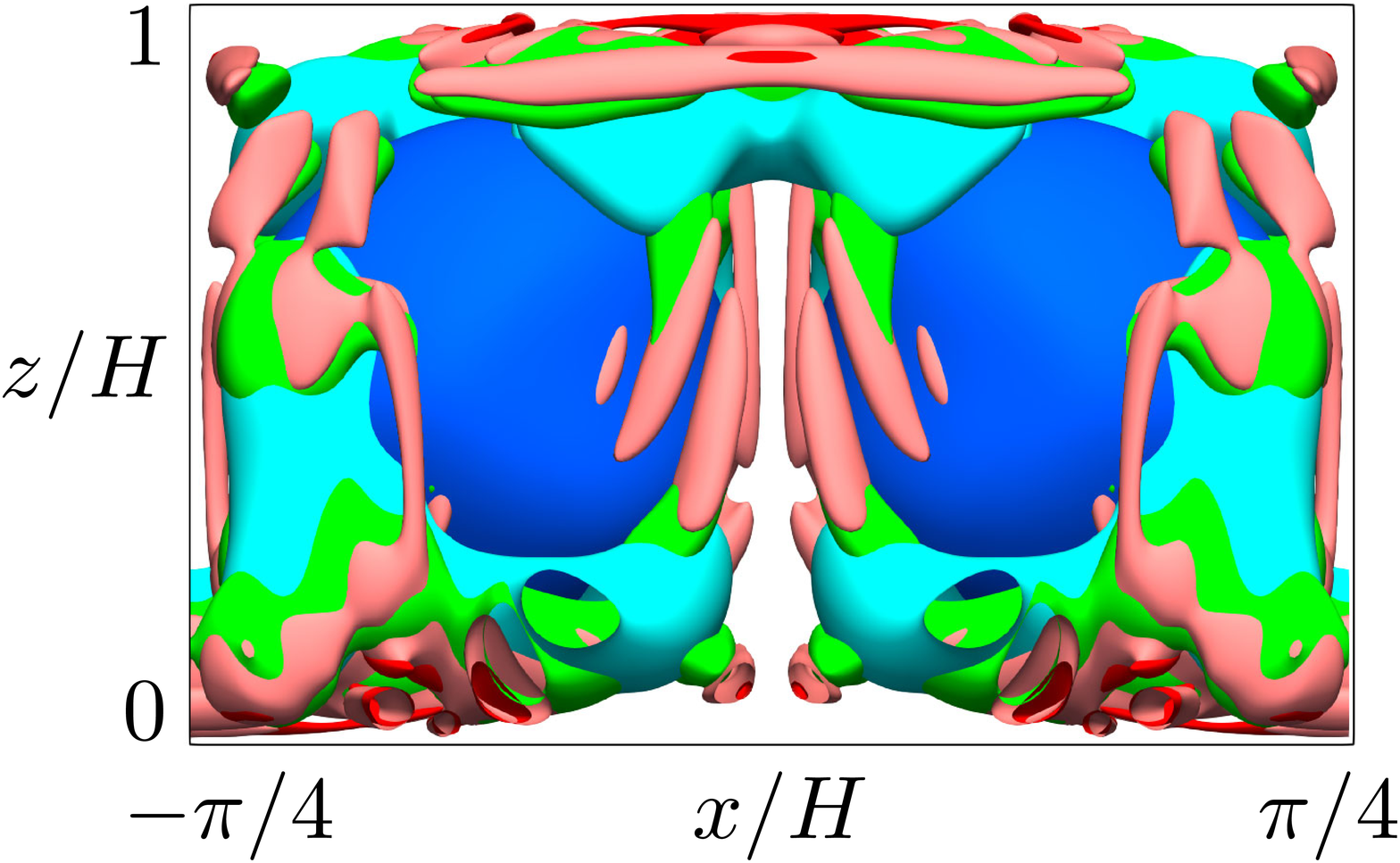}
	\end{minipage}
\caption{Hierarchical vortex structures visualised by coarse graining with Gaussian low-pass filter.
(\textit{a}) The yellow and red objects are the isosurfaces of the non-filtered $T/\Delta T=0.6$ and $Q/(\kappa^2/H^4)=2\times10^8$, respectively.
(\textit{b-h}) The vortex structures are visualised by the isosurfaces of $Q/(\kappa^2/H^4)$ of the filtered velocity field with a filter widths of $\sigma=H(=2L/\pi)$ (blue), $\sigma=L/4$ (light blue), $\sigma=L/8$ (green), $\sigma=L/16$ (light red), $\sigma=L/32$ (red), and they are superposed in (\textit{g,h}).
The isosurface levels are (blue) $5\times10^5$, (light blue) $4\times10^6$, (green) $1.2\times10^7$, (light red) $3\times10^7$ and (red) $1.6\times10^8$.
\label{fig:coarse}}
\end{figure}

The developed turbulence organises hierarchical coherent vortex structures of various scales \citep{Goto2017,Motoori2019}; however, it is difficult to identify large- and intermediate-scale structures.
The smallest-scale vortex structures can still be extracted by employing the isosurface of $Q$, as shown in figure \ref{fig:structure}.
To examine the hierarchy of multi-scale vortices in the 3D steady solution, we consider coarse graining the velocity field $\textit{\textbf{u}}$.
The coarse-grained velocity field $\textit{\textbf{u}}^{*}$ is obtained by the Gaussian low-pass filter \citep{Duran2016,Motoori2019} as follows
\begin{eqnarray}
\displaystyle
\textit{\textbf{u}}^{*}(\textit{\textbf{x}})=\int_{V} a\cdot \textit{\textbf{u}}(\textit{\textbf{x}}')\exp{\left\{ -{\left( \frac{\pi\Delta r}{\sigma} \right)}^2 \right\}} {\rm d}\textit{\textbf{x}}',
\end{eqnarray}
where $\Delta r=|\mbox{\boldmath$x$}'-\mbox{\boldmath$x$}|$, $\sigma$ is the filter width and $a$ is a constant such that the integral of the kernel over the control volume $V$ is unity.
In the wall-normal direction, the Gaussian filter is applied by reflecting it at the wall \citep{Duran2016}.
Figure \ref{fig:coarse} shows hierarchical vortex structures in the 3D steady solution at $Ra=2.6\times10^7$.
Non-filtered structures are shown in figure \ref{fig:coarse}(\textit{a}), and the isosurfaces of $Q$ of the filtered velocity ${\textit{\textbf{u}}}^{*}$ with $\sigma=H(=2L/\pi),L/2,L/4,L/8$ and $L/16$ are displayed in figure \ref{fig:coarse}(\textit{b}-\textit{f}), respectively.
The blue objects in figure \ref{fig:coarse}(\textit{b}) are the largest-scale structures corresponding to the large-scale convection, whereas the red ones in figure \ref{fig:coarse}(\textit{f}) are the smallest-scale structures of size $\sigma/2=L/32\approx2\delta$, which coincides with the size of vortices observed in the non-filtered field (figure \ref{fig:coarse}\textit{a}).
The light blue, green and light red objects in figure \ref{fig:coarse}(\textit{c,d,e}) illustrate the intermediate-scale vortex structures with eight, four and two times the size of the smallest-scale vortices, respectively.
The smaller-scale vortex structures exist closer to the wall, while the intermediate-scale ones are observed in the bulk region in figure \ref{fig:coarse}(\textit{d,e}).

\begin{figure}
\centering
	\begin{minipage}{.7\linewidth}
	(\textit{a})\\
	\includegraphics[clip,width=\linewidth]{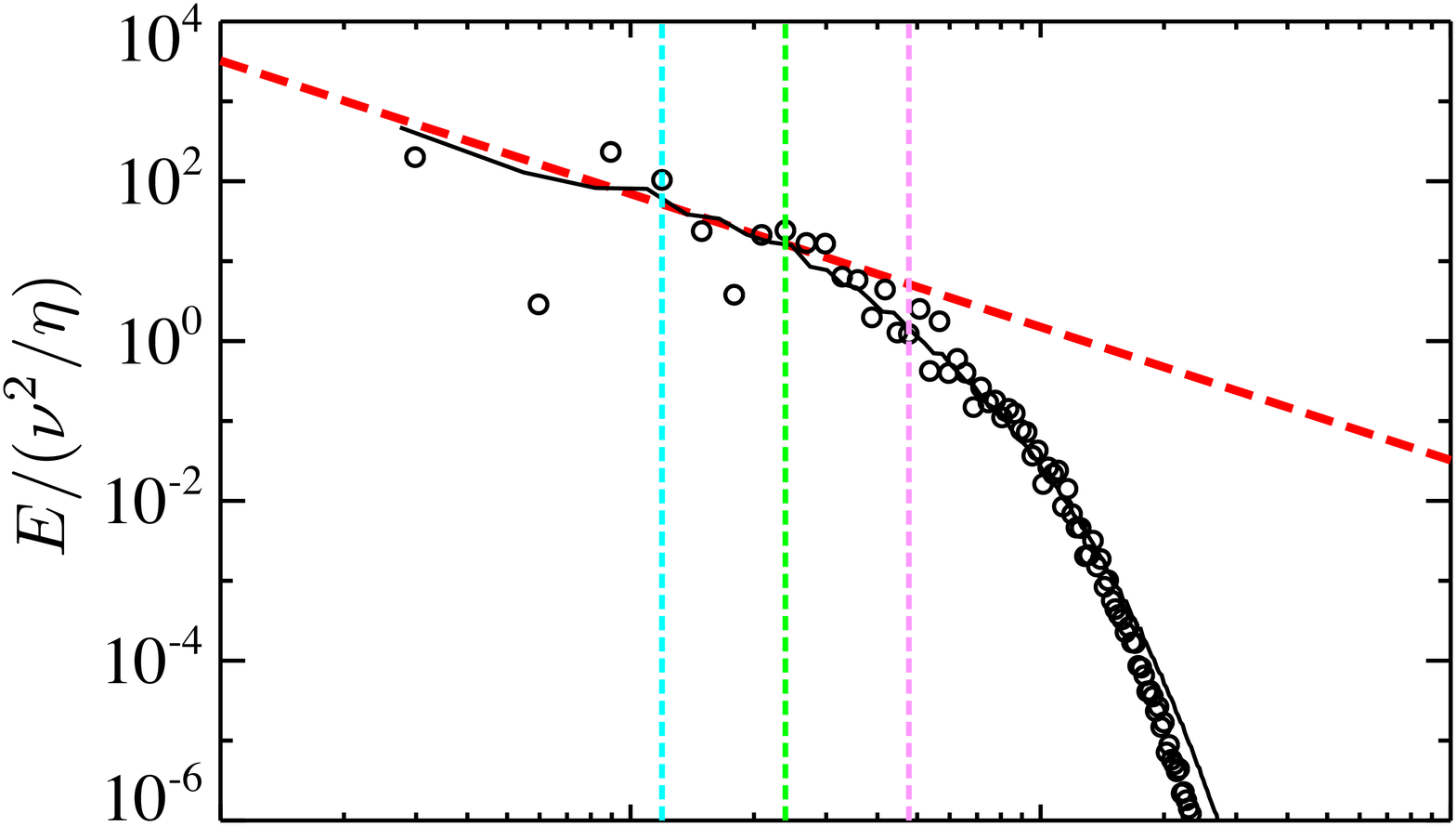}
	\end{minipage}
	
	\begin{minipage}{.7\linewidth}
	(\textit{b})\\
	\includegraphics[clip,width=\linewidth]{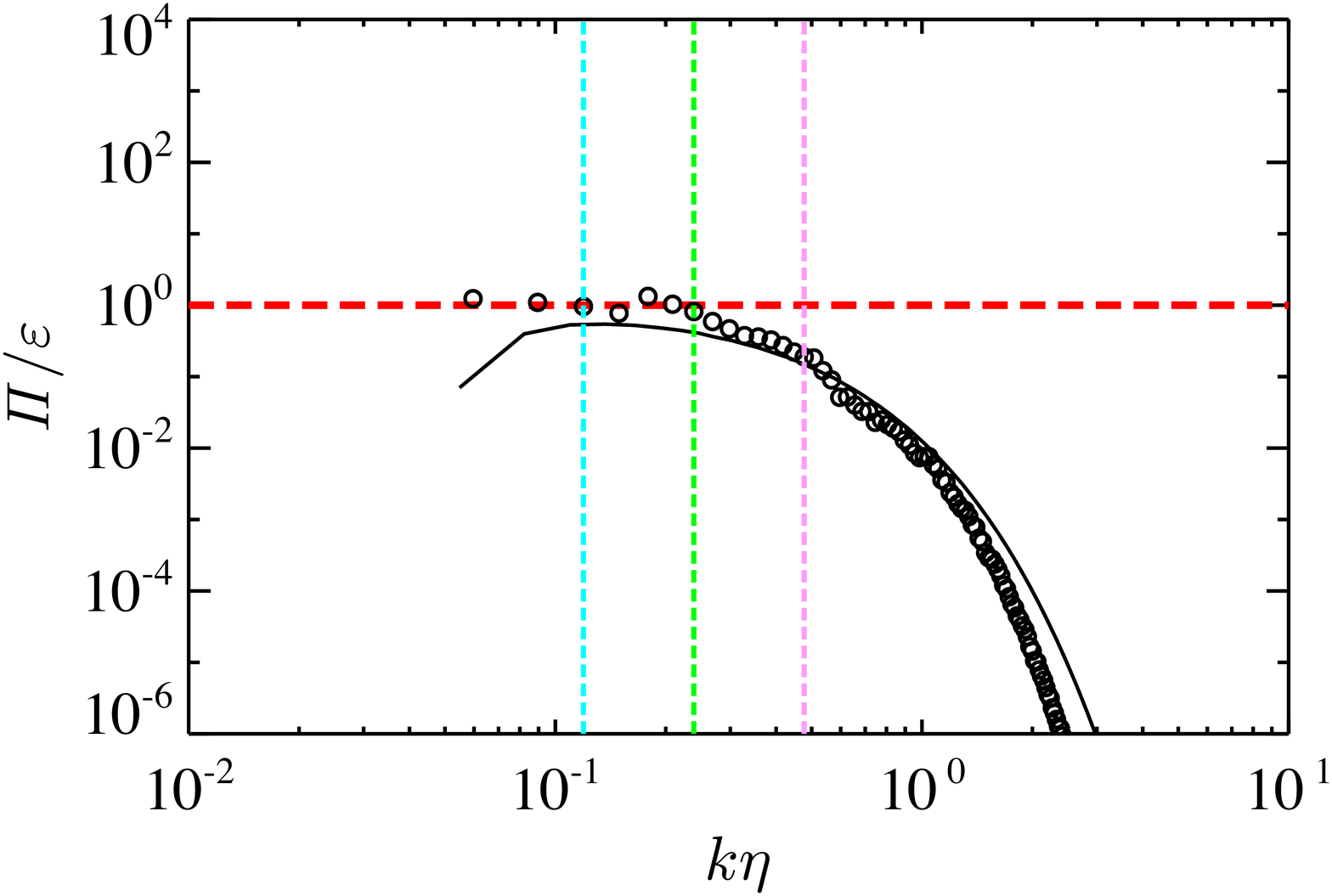}
        \end{minipage}
\caption{(\textit{a}) Energy spectrum $E$ and (\textit{b}) energy flux $\varPi$ at the centre of the fluid layer, $z=H/2$, in the 3D steady solution (circles) and the turbulent state (lines) at $Ra=2.6\times10^7$.
The lateral and longitudinal axes are normalised by the kinematic viscosity $\nu$ and the energy dissipation rate $\varepsilon$ at $z=H/2$, where $\eta={(\nu^{3}/\varepsilon)}^{1/4}$ is the Kolmogorov micro-scale length.
The red dashed lines represent $E=1.5\varepsilon^{2/3}k^{-5/3}$ and $\varPi/\varepsilon=1$, respectively.
The light blue, green and light red colours indicate $k=2\pi/(L/4)$, $2\pi/(L/8)$ and $2\pi/(L/16)$, respectively, normalised with $\eta$ in the 3D steady solution, corresponding to the intermediate-scale structures shown in figure \ref{fig:coarse}.
\label{fig:spectra}}
\end{figure}

Figure \ref{fig:coarse}(\textit{g,h}) shows the superposed structures, and from their spatial distribution it is conjectured that the bulk flow is composed of multi-scale coherent structures.
Figure \ref{fig:spectra}(\textit{a}) shows the energy spectrum $E(k,z)$ of the 3D steady solution at the centre of the fluid layer, $z=H/2$, and the corresponding turbulence spectrum at $Ra=2.6\times10^{7}$.
$E(k,z)$ is defined as
\begin{eqnarray}
\displaystyle
E(k,z)=\frac{L}{2\pi}\sum_{k-\frac{\Delta k}{2}<|\mbox{\boldmath$k$}_{\rm 2D}|<k+\frac{\Delta k}{2}}\frac{1}{2}{\left< {|\widetilde{\mbox{\boldmath$u$}}(\mbox{\boldmath$k$}_{\rm 2D},z)|}^{2} \right>}_{t},
\end{eqnarray}
where $\widetilde{(\cdot)}$ indicates the Fourier coefficients in the periodic ($x$- and $y$-) directions, and ${\langle\cdot \rangle}_{t}$ represents the time average.
$\mbox{\boldmath$k$}_{\rm 2D}=(k_{x},k_{y})$ and $k=|\mbox{\boldmath$k$}_{\rm 2D}|$ are the wavenumber vector and its magnitude, respectively, and $\Delta k=2\pi/L$.
The lateral and longitudinal axes are normalised by the kinematic viscosity $\nu$ and the energy dissipation rate
\begin{eqnarray}
\displaystyle
\varepsilon(z)=\frac{\nu}{2}{\left< {\left( \frac{\partial u_{i}}{\partial x_{j}}+\frac{\partial u_{j}}{\partial x_{i}} \right)}^{2} \right>}_{xyt}.
\end{eqnarray}
$\eta={(\nu^3/\varepsilon)}^{1/4}$ is the Kolmogorov micro-scale length.
The spectra of the 3D steady solution and turbulent state are in good agreement with the high-wavenumber $k\eta\gtrsim10^{0}$.
Furthermore, in the wavenumber band of $2\pi/(L/4)\lesssim k\eta\lesssim2\pi/(L/16)$, corresponding to the intermediate-scale range, the energy spectrum follows Kolmogorov's $-5/3$ power law, $E=C_{K}\varepsilon^{2/3}k^{-5/3}$ \citep{Kolmogorov1941}, with the constant, $C_{K}\approx1.5$, which is consistent with that in the inertial subrange of high-Reynolds-number turbulence \citep{Sreenivasan1995,Ishihara2016}.

In figure \ref{fig:spectra}(\textit{b}), we show the energy flux in the wavenumber space, $\varPi(k,z)$ \citep{Mizuno2016}, defined as
\begin{eqnarray}
&\displaystyle
\varPi(k,z)=\sum_{k'\ge k}\sum_{k-\frac{\Delta k}{2}<|\textit{\textbf{k}}_{\rm 2D}|<k+\frac{\Delta k}{2}}T^{s}(\textit{\textbf{k}}_{\rm 2D},z),&\\
&\displaystyle
T^{s}(\textit{\textbf{k}}_{\rm 2D},z)=\Re\left[ {\left< \partial_{j}\widetilde{u}_{i}{(\widetilde{u_{i}u_{j}})}^{\dagger} \right>}_{t}-\frac{1}{2}\frac{\partial {\left< \widetilde{u}_{j}{(\widetilde{u_{j}w})}^{\dagger} \right>}_{t}}{\partial z} \right],&
\end{eqnarray}
where $(\partial_{1},\partial_{2},\partial_{3})=({\rm i}k_{x},{\rm i}k_{y},\partial/\partial z)$ and $\dagger$ denotes the complex conjugate.
$T^{s}(\mbox{\boldmath$k$}_{\rm 2D},z)$ represents the energy transfer between the Fourier modes, and the sum of all spectral components does not contribute to the total energy budget, i.e., $\sum_{\mbox{\boldmath$k$}_{\rm 2D}}T^{s}(\textit{\textbf{k}}_{\rm 2D},z)=0$.
In the intermediate-scale range, the energy flux exhibits positive values, that is, the energy transfer from large to small scale, and it scales with the same order of energy dissipation rate.

\section{Summary and conclusions}\label{sec:summary}
We have discovered a three-dimensional steady solution to the Boussinesq equations that exhibits scaling ($Nu\sim Ra^{0.31}$) and multi-scale coherent structures, which are similar to those observed in turbulent Rayleigh--B\'enard convection.
The invariant solution bifurcates from the conduction state at $Ra\sim10^3$, and it has been tracked up to $Ra\sim10^7$ by using the Newton--Krylov iteration.
The horizontal-averaged temperature and the RMS of the temperature and velocity fluctuations are in good agreement with the horizontal and temporal averages for the turbulent states.
In the near-wall region, smaller-scale thermal plumes are generated with an increase in $Ra$.
The size of the thermal coherent structures and relevant vortices is comparable with the thermal conduction layer thickness $\delta/H=1/(2Nu)$, and the RMS vertical velocity at $z/\delta\sim1$ scales with the velocity scale $Ra^{1/3}\kappa/H$, corresponding to $Nu\sim Ra^{1/3}$.
On the other hand, in the bulk region, the flow consists of hierarchical multi-scale vortices.
We have extracted the large- and intermediate-scale vortex structures by employing the coarse-graining method.
The ratio of the largest to the smallest length scales in the 3D steady solution at $Ra=2.6\times10^{7}$ is approximately $20$.
The energy spectrum at the centre of the fluid layer shows good agreement with that of the turbulent state.
In the intermediate-scale range, the spectrum follows $E=1.5\varepsilon^{2/3}k^{-5/3}$, which is commonly observed in the inertial subrange of the developed turbulence.
Furthermore, energy is transferred from large to small scales in the wavenumber space, and the energy flux balances the energy dissipation rate, in accordance with the Kolmogorov--Obukhov energy cascade view.

Recently, \cite{Veen2019} have found a time-periodic solution that reproduces inertial range dynamics in a triply periodic turbulence driven by a constant body force of the Taylor--Green type.
They have obtained the invariant solution by applying large eddy simulation based on the Smagorinsky-type eddy-viscosity model.
By introducing the buoyant force, meanwhile, we have succeeded in finding a multi-scale solution of the full incompressible Navier--Stokes equation without any empirical models.
We believe that the current work and approaches based on multi-scale invariant solutions will trigger significant advances in the theoretical understanding and deductive modelling of coherent structures and energy transfer mechanisms in developed turbulence.

\section*{Acknowledgements}
This work was supported by the Japanese Society for Promotion of Science (JSPS) KAKENHI (Grant Numbers 19K14889 and 18H01370).
In this research work we used the supercomputer of ACCMS, Kyoto University.
This work was supported by NIFS Collaboration Research program (NIFS19KNSS124).

\appendix

\begin{figure}
\centering
	\begin{minipage}{.6\linewidth}
	(\textit{a})\\
	\includegraphics[clip,width=\linewidth]{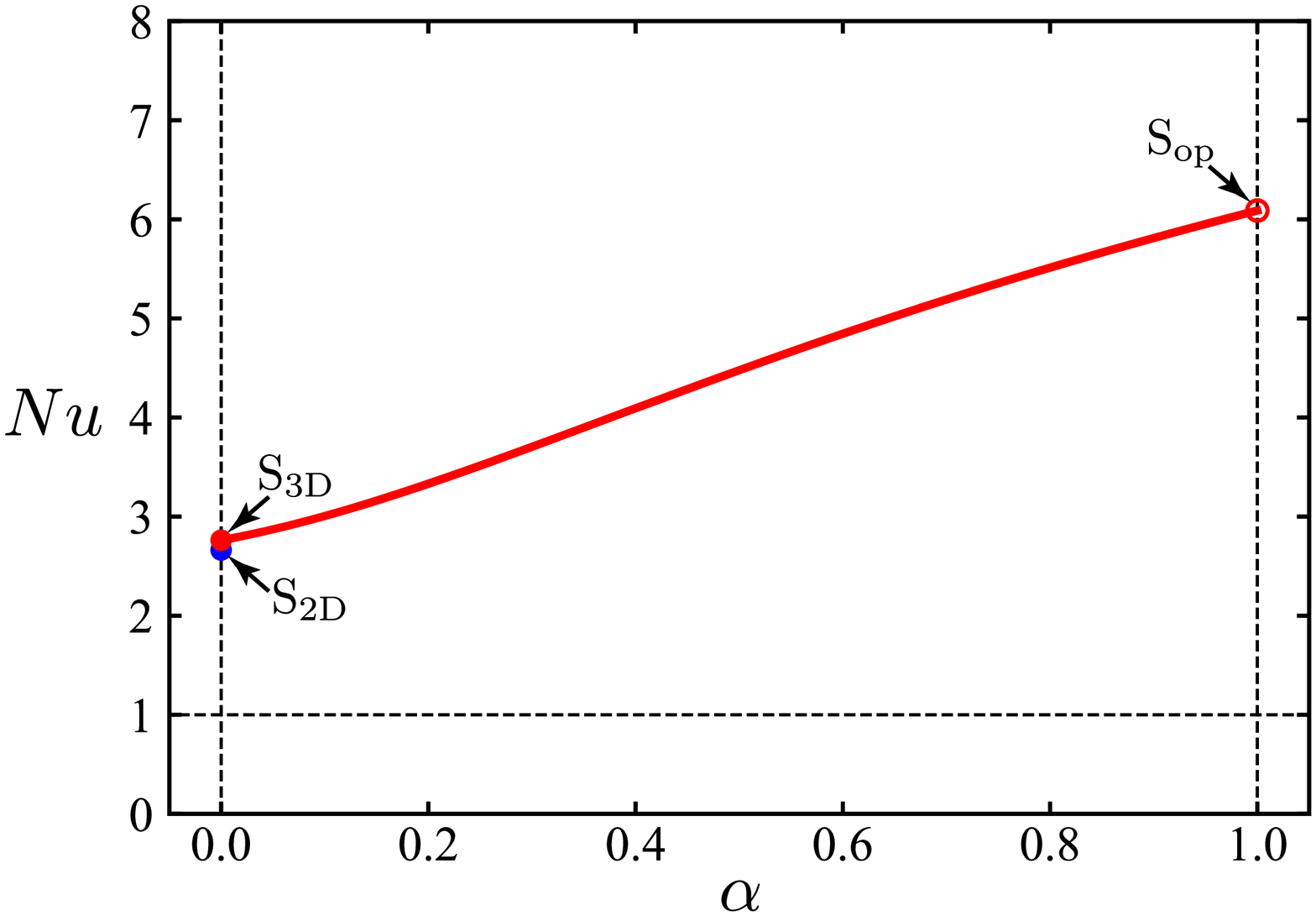}
	\end{minipage}
	
        \vspace{1em}
	\begin{minipage}{.32\linewidth}
	(\textit{b})\\
	\includegraphics[clip,width=\linewidth]{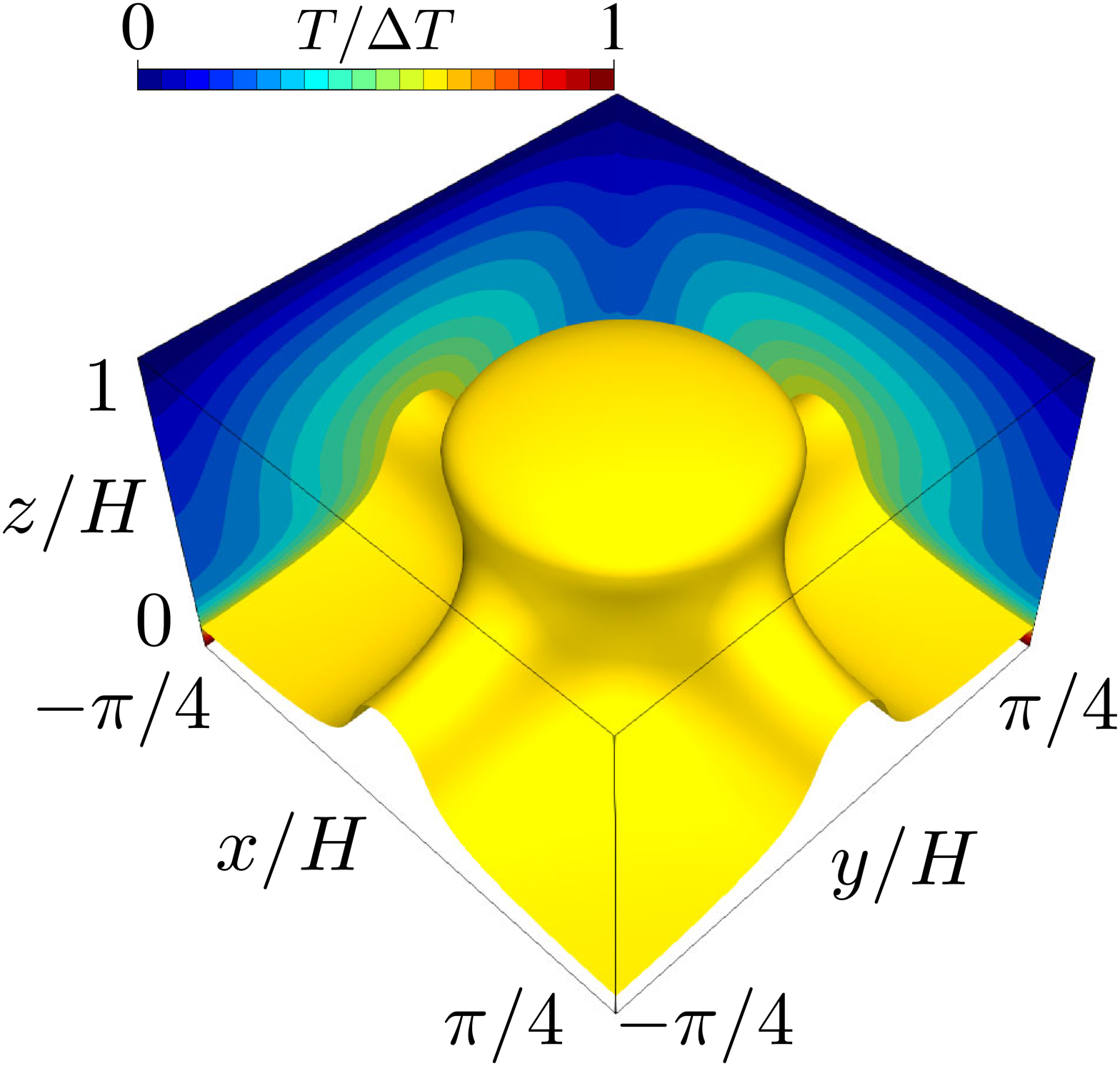}
	\end{minipage}
	\begin{minipage}{.32\linewidth}
	(\textit{c})\\
	\includegraphics[clip,width=\linewidth]{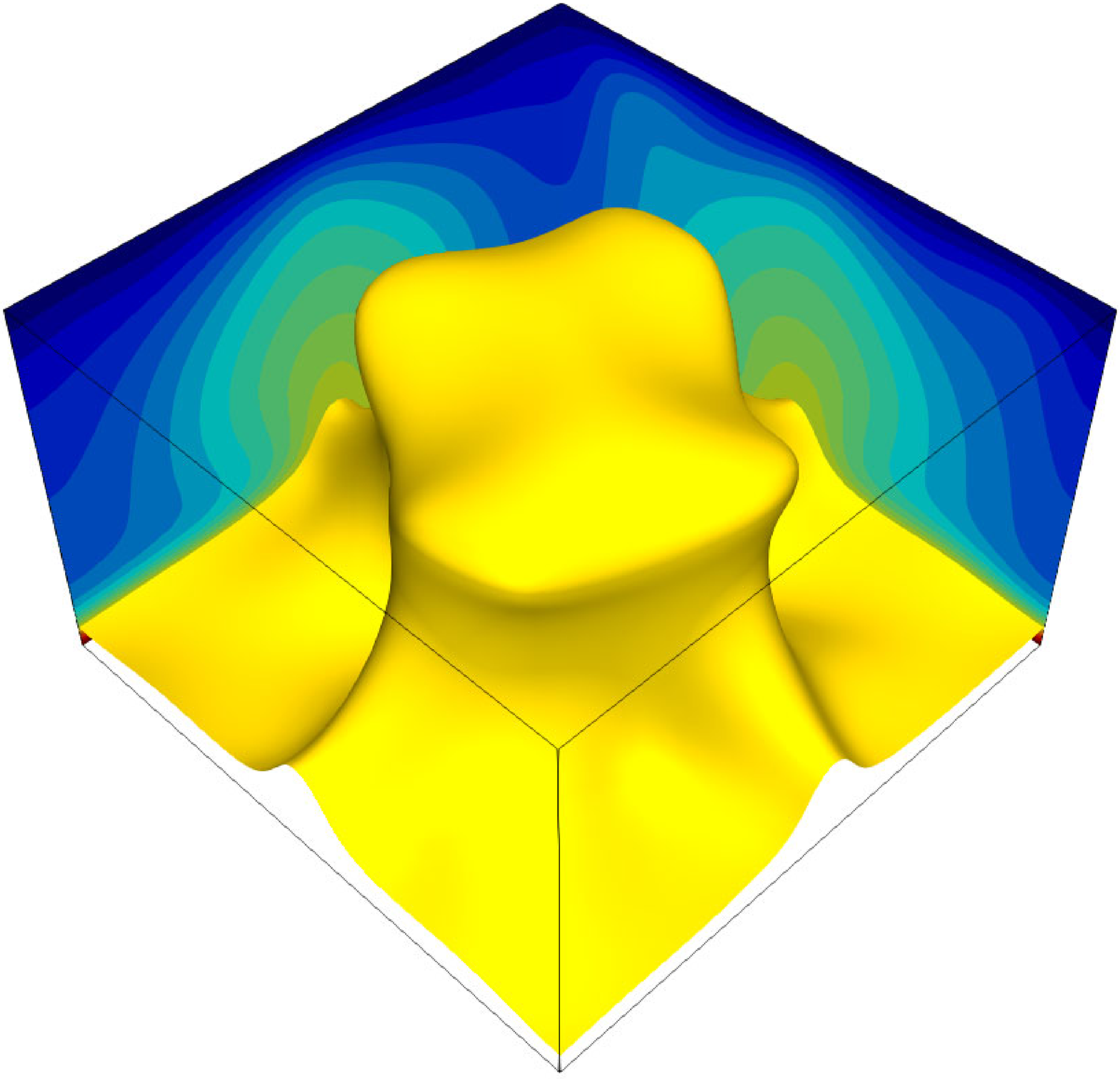}
        \end{minipage}
	\begin{minipage}{.32\linewidth}
	(\textit{d})\\
	\includegraphics[clip,width=\linewidth]{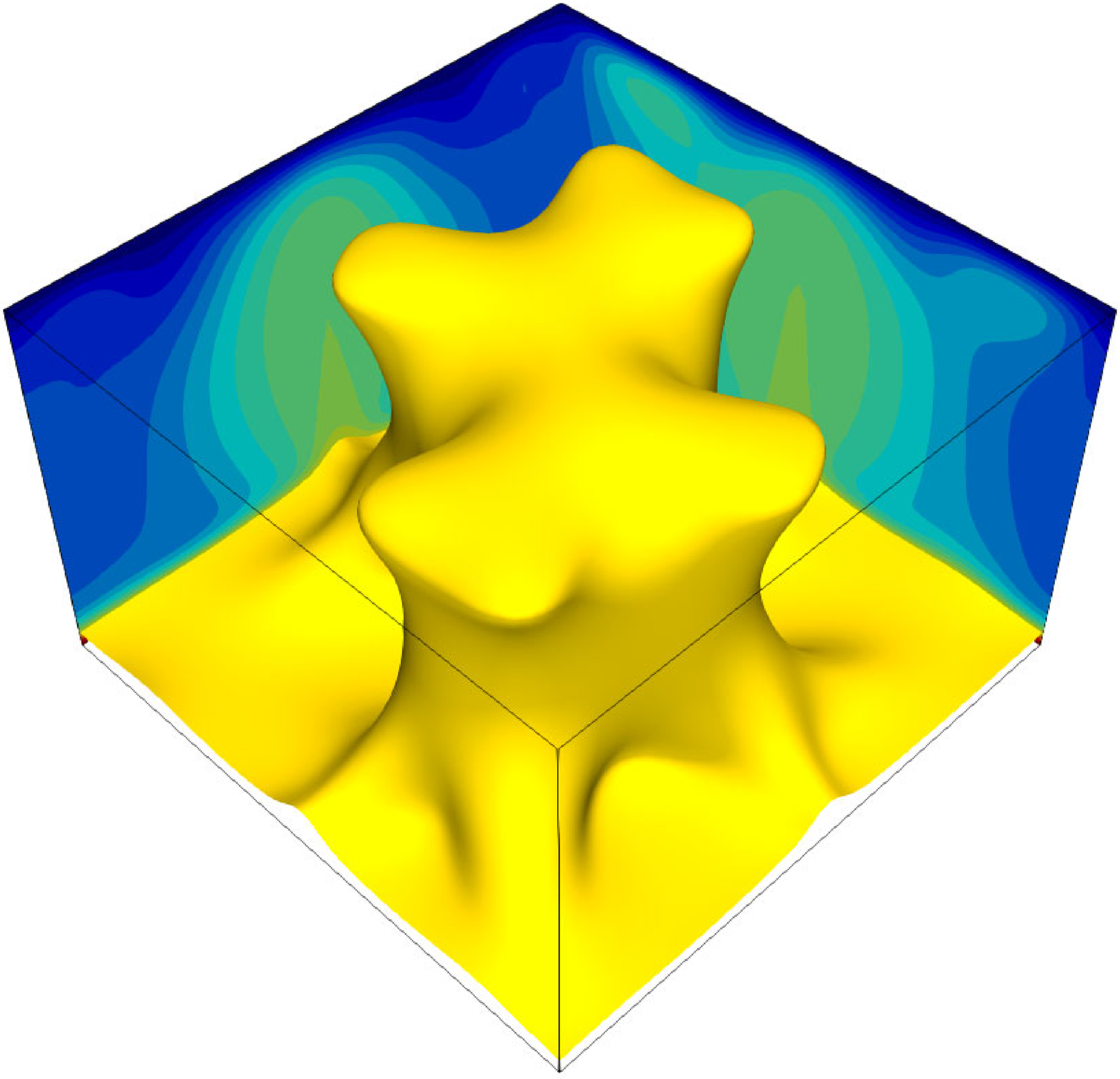}
	\end{minipage}
\caption{Homotopy from the wall-to-wall optimal transport solution at $Pe=508$ \citep[from][]{Motoki2018b} to the present 3D steady solution for a fixed $Ra=10^{4}$ and $Pr=1$.
(\textit{a}) Nusselt number $Nu$ as a function of the homotopy parameter $\alpha$.
The red open circle shows the optimal solution ${\rm S}_{\rm op}$ of the Euler--Lagrange equations for the wall-to-wall optimal transport problem, and the red and blue filled circles represent the 3D steady solution ${\rm S}_{\rm 3D}$ and the 2D steady solution ${\rm S}_{\rm 2D}$ of the Boussinesq equations, respectively.
(\textit{b-d}) Isosurfaces of temperature $T/\Delta T=0.6$ at (\textit{b}) $\alpha=0$, (\textit{c}) $\alpha=0.3$ and (\textit{d}) $\alpha=1.0$.
The contours represent the temperature $T$ on the planes $x/H=-\pi/4$ and $y/H=\pi/4$.
The numerical computation is carried out on $64^{3}$ grid points.
\label{fig:homotopy}}
\end{figure}

\section{Homotopy from wall-to-wall optimal transport solution}\label{sec:homotopy}
The wall-to-wall optimal transport problem \citep{Hassanzadeh2014,Motoki2018b,Souza2020} involves maximising the heat flux between two parallel plates with a constant temperature difference, under the constraint of fixed total enstrophy, which is written as
\begin{eqnarray}
&{\ }&\displaystyle {\rm Maximise}\hspace{1em}Nu=1+{\left< w\theta \right>}_{xyz} \nonumber\\
&{\ }&\displaystyle {\rm subject{\ }to}\hspace{1em}\nabla\cdot\mbox{\boldmath$u$}=0, \nonumber \\
&{\ }&\displaystyle \hspace{5em}(\textit{\textbf{u}}\cdot\nabla)\theta+w=\kappa\nabla^{2}\theta, \nonumber \\
&{\ }&\displaystyle \hspace{5em}Pe=\frac{{\left< {|\nabla\cdot\mbox{\boldmath$u$}|}^{2} \right>}_{xyz}^{1/2}H^{2}}{\kappa}={\rm const.}, \nonumber \\
&{\ }&\displaystyle \hspace{5em}{\rm and{\ }the{\ }boundary{\ }conditions},
\end{eqnarray}
where $\theta=T-(1-z)$ is the temperature fluctuation about a conduction state.
The constraint optimisation is relevant to the maximisation of the objective functional
\begin{eqnarray}
\displaystyle
\mathcal{F}'={\left< w'\theta'-\phi'(\mbox{\boldmath$x$}')[(\mbox{\boldmath$u$}'\cdot\nabla')\theta'+w'-{\nabla'}^{2}\theta']
+\psi'(\mbox{\boldmath$x$}')(\nabla'\cdot\mbox{\boldmath$u$}')+\frac{\mu'}{2}(Pe^{2}-{|\nabla'\mbox{\boldmath$u$}'|}^{2}) \right>}_{xyz}, \nonumber\\
\end{eqnarray}
where $\phi'(\mbox{\boldmath$x$}')$, $\psi'(\mbox{\boldmath$x$}')$ and $\mu'$ are Lagrange multipliers, and prime $(\cdot)'$ represents a non-dimensional variable based on $H$, $\Delta T$, $\kappa$ and $\rho$.
The Euler--Lagrange equations are
\begin{eqnarray}
\label{eq:el1}
\displaystyle
\frac{\delta \mathcal{F}'}{\delta \mbox{\boldmath$u$}'}&\equiv&
-\nabla' \psi'+\theta'\nabla'\phi'+\mu'{\nabla'}^{2}\mbox{\boldmath$u$}'+(\theta'+\phi')\mbox{\boldmath$e$}_{z}=\mbox{\boldmath$0$},\\
\label{eq:el2}
\displaystyle
\frac{\delta \mathcal{F}'}{\delta \theta'}&\equiv&
(\mbox{\boldmath$u$}'\cdot\nabla')\phi'+w'+{\nabla'}^{2}\phi'=0, \\
\label{eq:el3}
\displaystyle
\frac{\delta \mathcal{F}'}{\delta \phi'}&\equiv&
-(\mbox{\boldmath$u$}'\cdot\nabla')\theta'+w'+{\nabla'}^{2}\theta'=0, \\
\label{eq:el4}
\displaystyle
\frac{\delta \mathcal{F}'}{\delta \psi'}&\equiv&
\nabla'\cdot\mbox{\boldmath$u$}'=0,\\
\label{eq:el5}
\displaystyle
\frac{\partial \mathcal{F}'}{\partial \mu'}&\equiv&
\frac{1}{2}{\left< Pe^{2}-|\nabla'\mbox{\boldmath$u$}'|^{2} \right>}_{xyz}=0.
\end{eqnarray}
In our previous work \citep{Motoki2018b}, we obtained the optimal state so as to satisfy the equations (\ref{eq:el1})--(\ref{eq:el5}).
Thus, to fulfil the Boussinesq equations, the optimal velocity and temperature field ($\mbox{\boldmath$u$}_{\rm op}'$, $\theta_{\rm op}'$) require an additional body force
\begin{eqnarray}
&{\ }&\displaystyle
\mbox{\boldmath$f$}'(\mbox{\boldmath$x$}')=-(\mbox{\boldmath$u$}_{\rm op}'\cdot\nabla')\mbox{\boldmath$u$}_{\rm op}'-\nabla' p_{\rm op}'+Pr{\nabla'}^{2}\mbox{\boldmath$u$}_{\rm op}'+PrRa(1-z'+\theta_{\rm op}')\textit{\textbf{e}}_{z},
\end{eqnarray}
which is different from the buoyant force, where $p_{\rm op}'$ is the pressure determined by the Poisson equation stemming from the Boussinesq equations.
We consider homotopy from the Euler--Lagrange system to the steady Boussinesq system
\begin{eqnarray}
&{\ }&\displaystyle
-(\mbox{\boldmath$u$}'\cdot\nabla')\mbox{\boldmath$u$}'-\nabla' p'+Pr{\nabla'}^{2}\mbox{\boldmath$u$}'+PrRa(1-z'+\theta')\textit{\textbf{e}}_{z}=\alpha \mbox{\boldmath$f$}',\\
&{\ }&\displaystyle
-(\mbox{\boldmath$u$}'\cdot\nabla')\theta'+w'+{\nabla'}^{2}\theta'=0,\\
&{\ }&\displaystyle
\nabla'\cdot\textit{\textbf{u}}'=0,
\end{eqnarray}
where $\alpha$ is a homotopy parameter.
For a fixed $Ra=10^{4}$, $Pr=1$ and $\mbox{\boldmath$f$}'$, we have tracked the solution from $\alpha=1$ to $0$ by using the Newton--Krylov method (figure \ref{fig:homotopy}).
The connected solution $S_{\rm 3D}$ is the present three-dimensional steady solution shown in \S \ref{sec:3d} and \S \ref{sec:hierarchy}.

\begin{figure}
\centering
	\begin{minipage}{.8\linewidth}
	\includegraphics[clip,width=\linewidth]{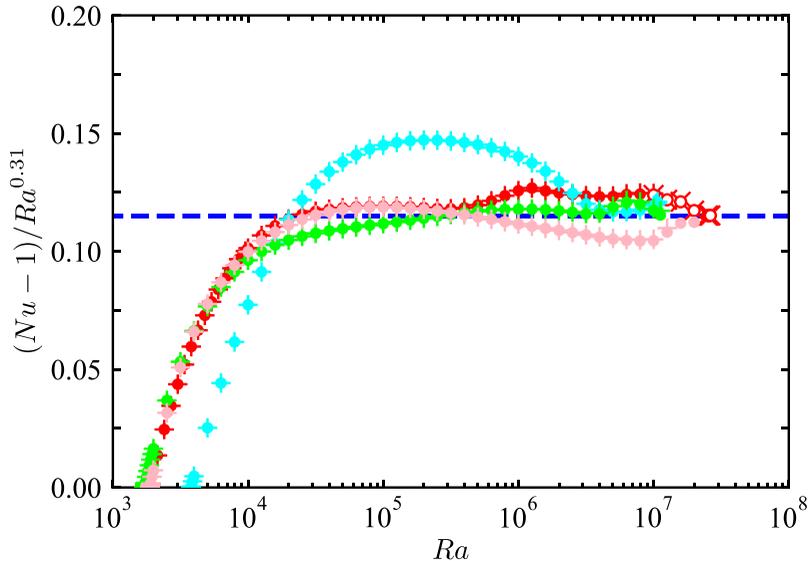}
	\end{minipage}
\caption{Nusselt number $Nu$ compensated by $Ra^{0.31}$ as a function of the Rayleigh number $Ra$ in the 3D steady solutions for the different horizontal period $L$ and Prandtl number $Pr$.
The green, red and light blue symbols show $L/H=2\pi/3.117$, $\pi/2$ and $1$, respectively, for $Pr=1$, and the light red symbols represent $L/H=\pi/2$ for $Pr=7$.
The blue dashed line indicates the optimal scaling in 2D steady solutions, $Nu-1=0.115Ra^{0.31}$ \citep{Waleffe2015,Sondak2015}.
The solutions have been obtained on grid points of  $+$, $(N_{x},N_{y},N_{z})=(64,64,64)$; $\bullet$, $(128,128,128)$; $\times$, $(192,192,128)$; $\circ$, $(256,256,256)$.
\label{fig:nu-ra_depend}}
\end{figure}

\begin{figure}
\centering
	\begin{minipage}{.32\linewidth}
	(\textit{a})\\
	\includegraphics[clip,width=\linewidth]{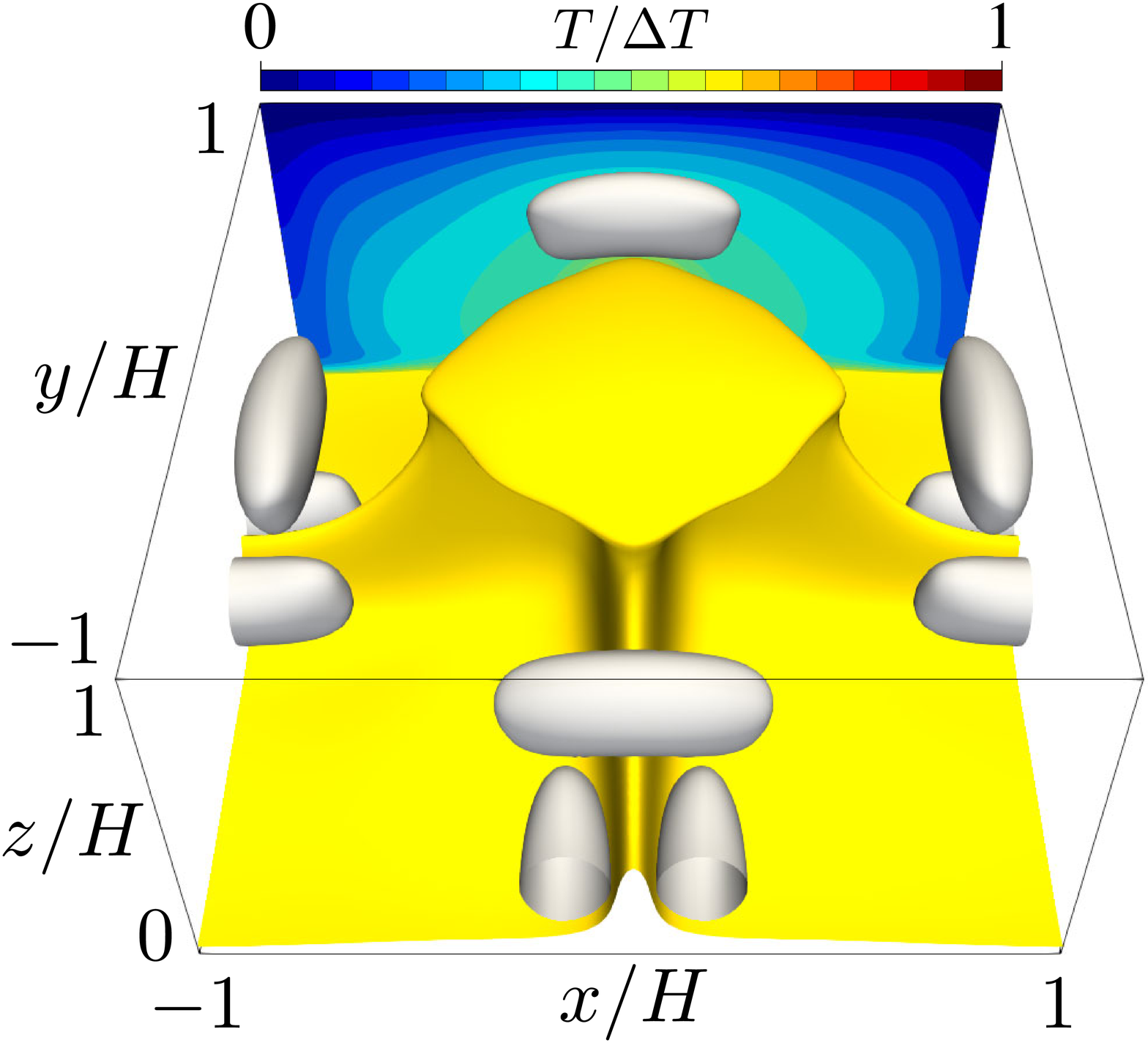}
	\end{minipage}
	\begin{minipage}{.32\linewidth}
	(\textit{b})\\
	\includegraphics[clip,width=\linewidth]{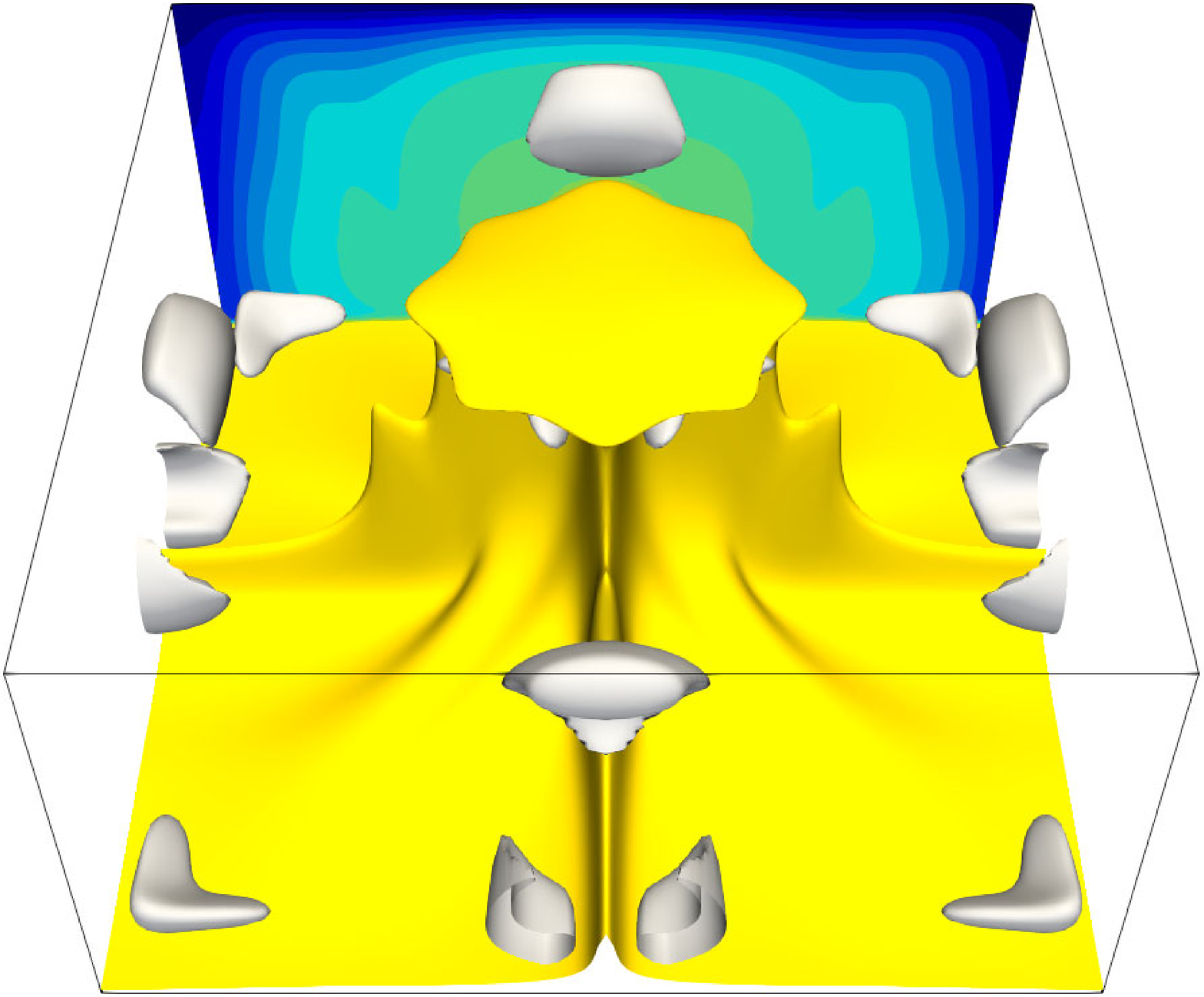}
        \end{minipage}
	\begin{minipage}{.32\linewidth}
	(\textit{c})\\
	\includegraphics[clip,width=\linewidth]{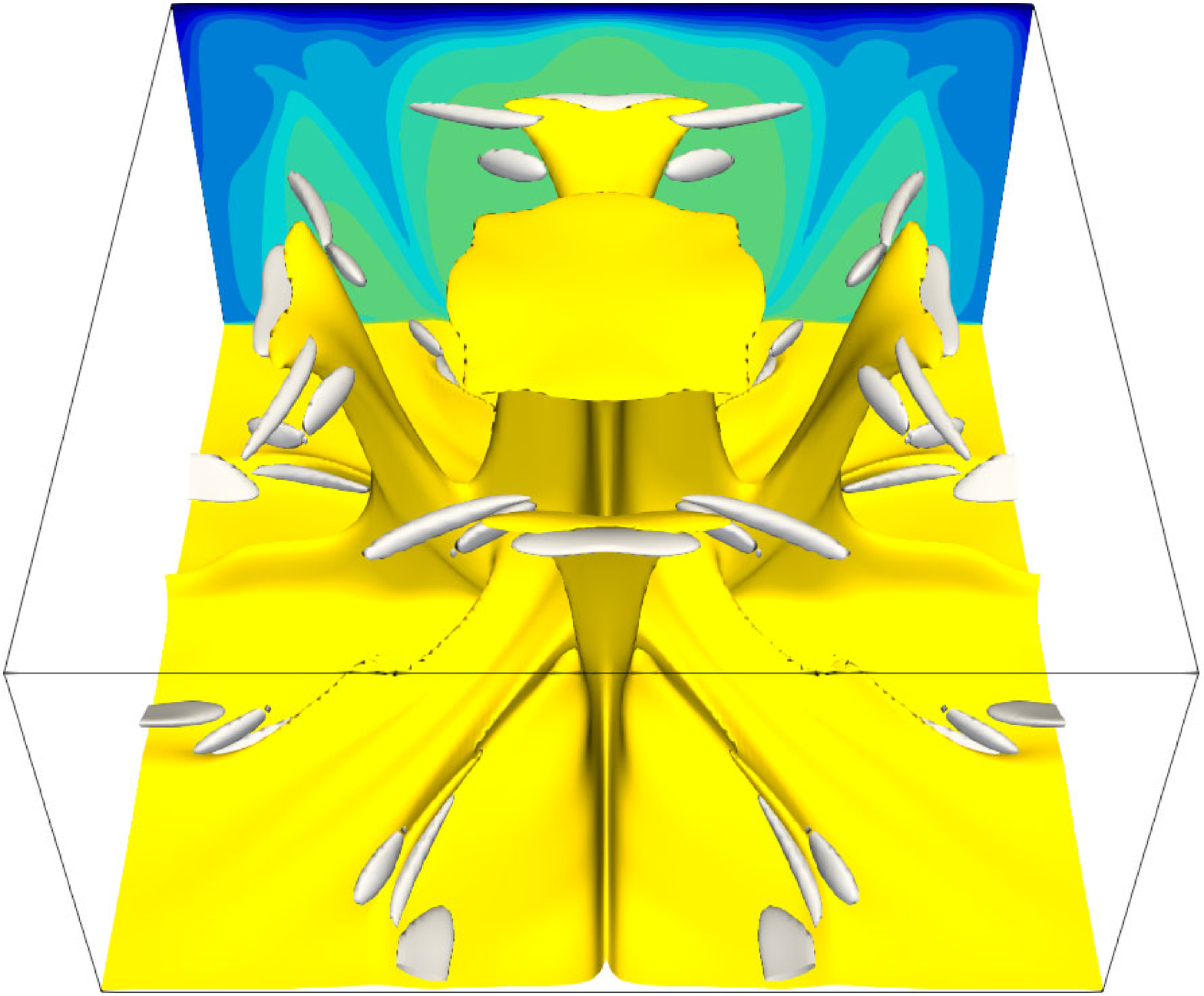}
	\end{minipage}
	
	\begin{minipage}{.32\linewidth}
	(\textit{d})\\
	\includegraphics[clip,width=\linewidth]{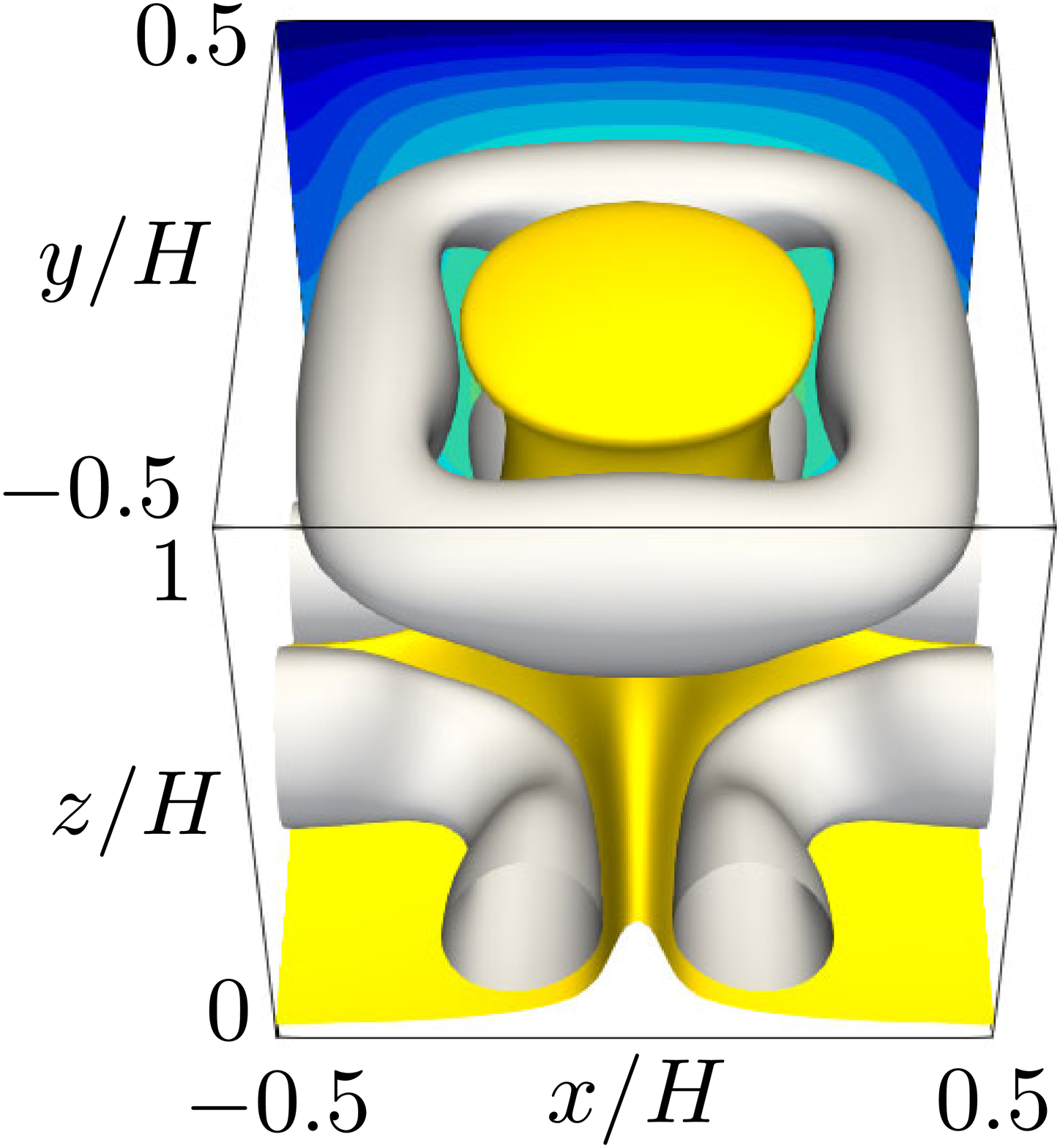}
	\end{minipage}
	\begin{minipage}{.32\linewidth}
	(\textit{e})\\
	\includegraphics[clip,width=\linewidth]{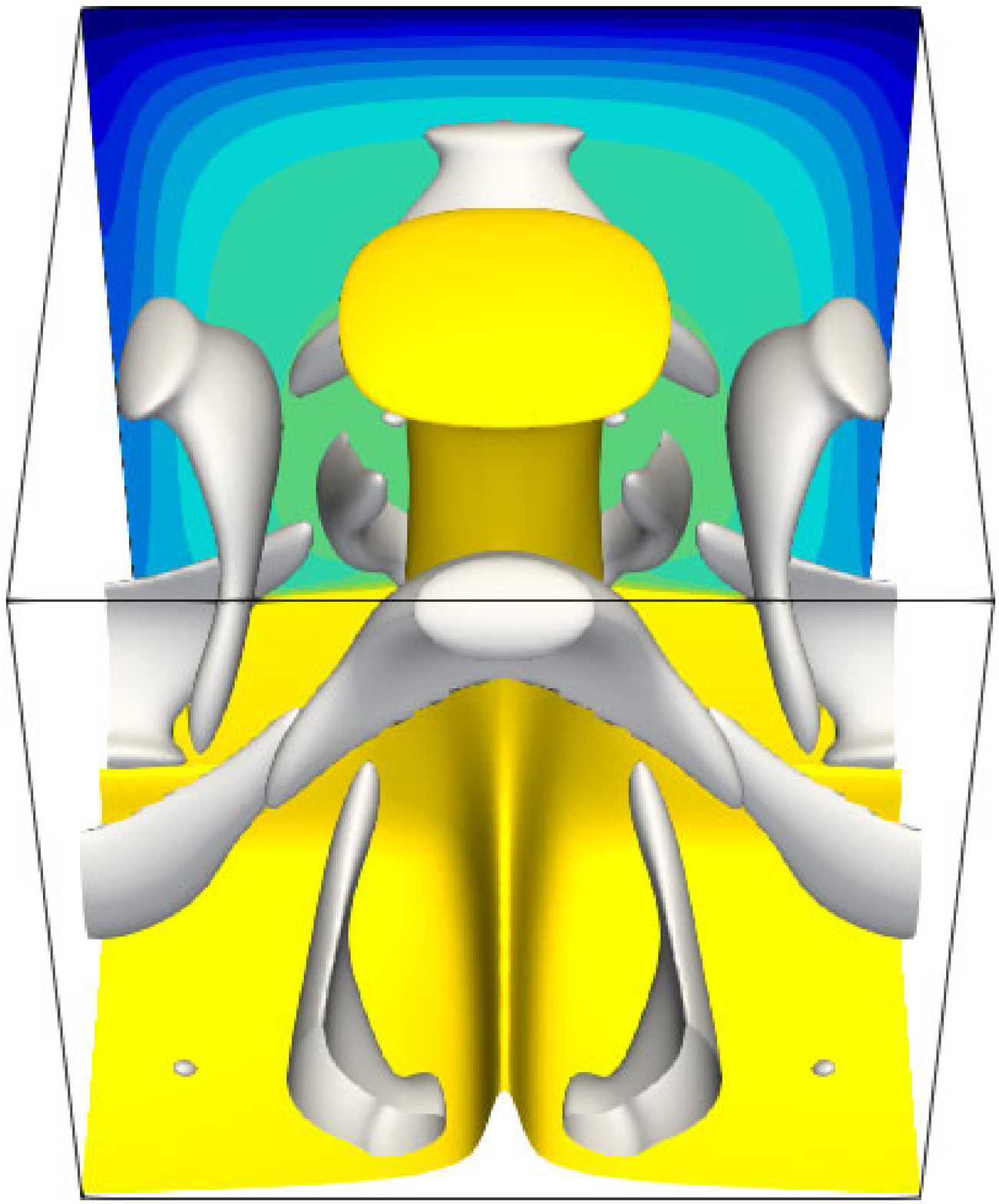}
        \end{minipage}
	\begin{minipage}{.32\linewidth}
	(\textit{f})\\
	\includegraphics[clip,width=\linewidth]{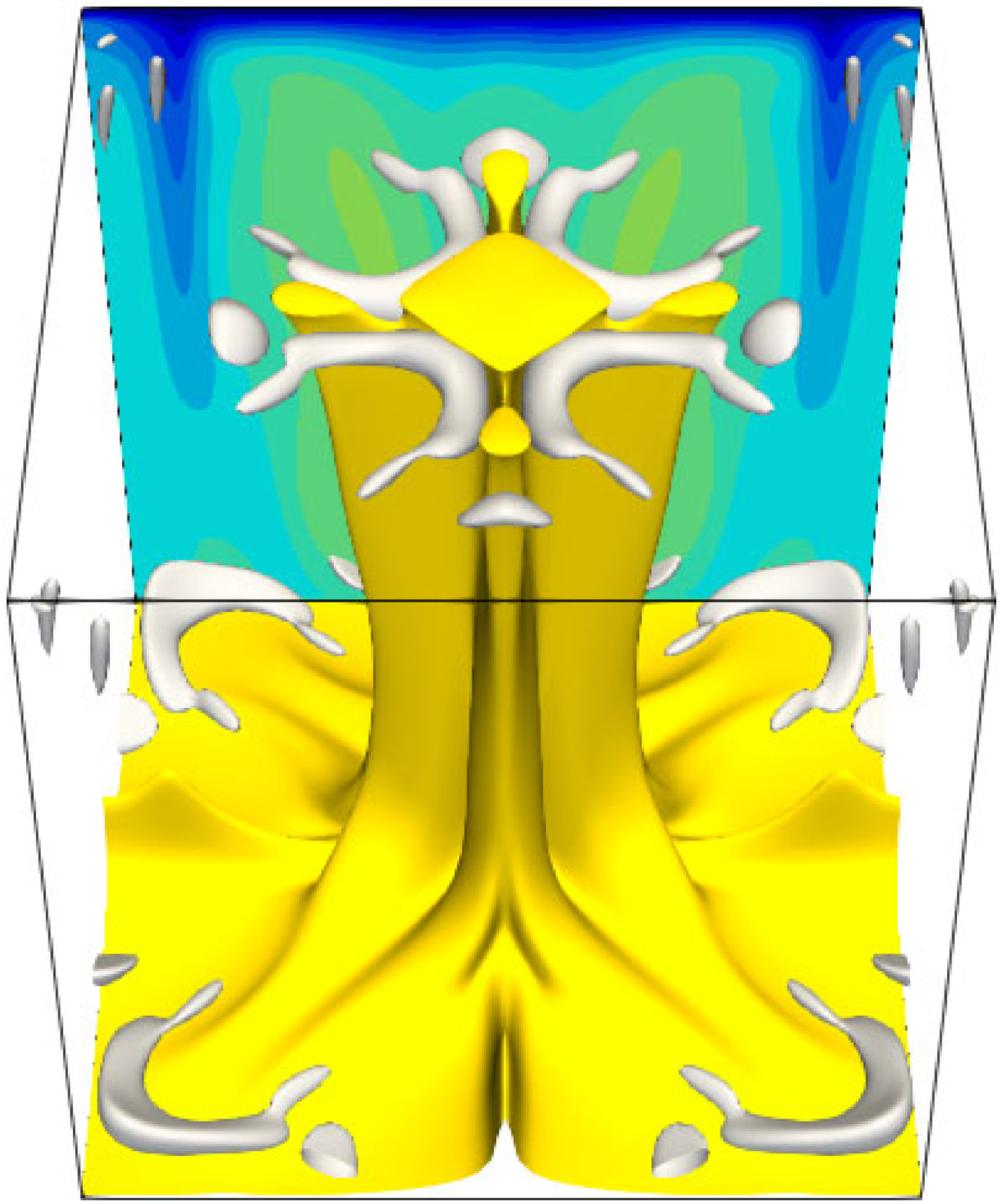}
	\end{minipage}
	
\caption{3D steady solutions in the domain with (\textit{a-c}) $L/H=2\pi/3.117\approx2.02$ and (\textit{d-f}) $L/H=1$ at (\textit{a,d}) $Ra=10^5$, (\textit{b,e}) $Ra=10^6$ and (\textit{c,f}) $Ra=10^7$ for $Pr=1$.
The yellow and grey objects show the isosurfaces of $T/\Delta T=0.6$ and (\textit{a,d}) $Q/(\kappa^2/H^4)=1.28\times10^5$, (\textit{b,e}) $Q/(\kappa^2/H^4)=2.4\times10^6$ and (\textit{c,f}) $Q/(\kappa^2/H^4)=8\times10^7$, respectively.
The contours represent $T$ in the plane (\textit{a-c}) $y/H=1$ and (\textit{d-f}) $y/H=0.5$.
\label{fig:aspect}}
\end{figure}

\begin{figure}
\centering
	\begin{minipage}{.32\linewidth}
	(\textit{a})\\
	\includegraphics[clip,width=\linewidth]{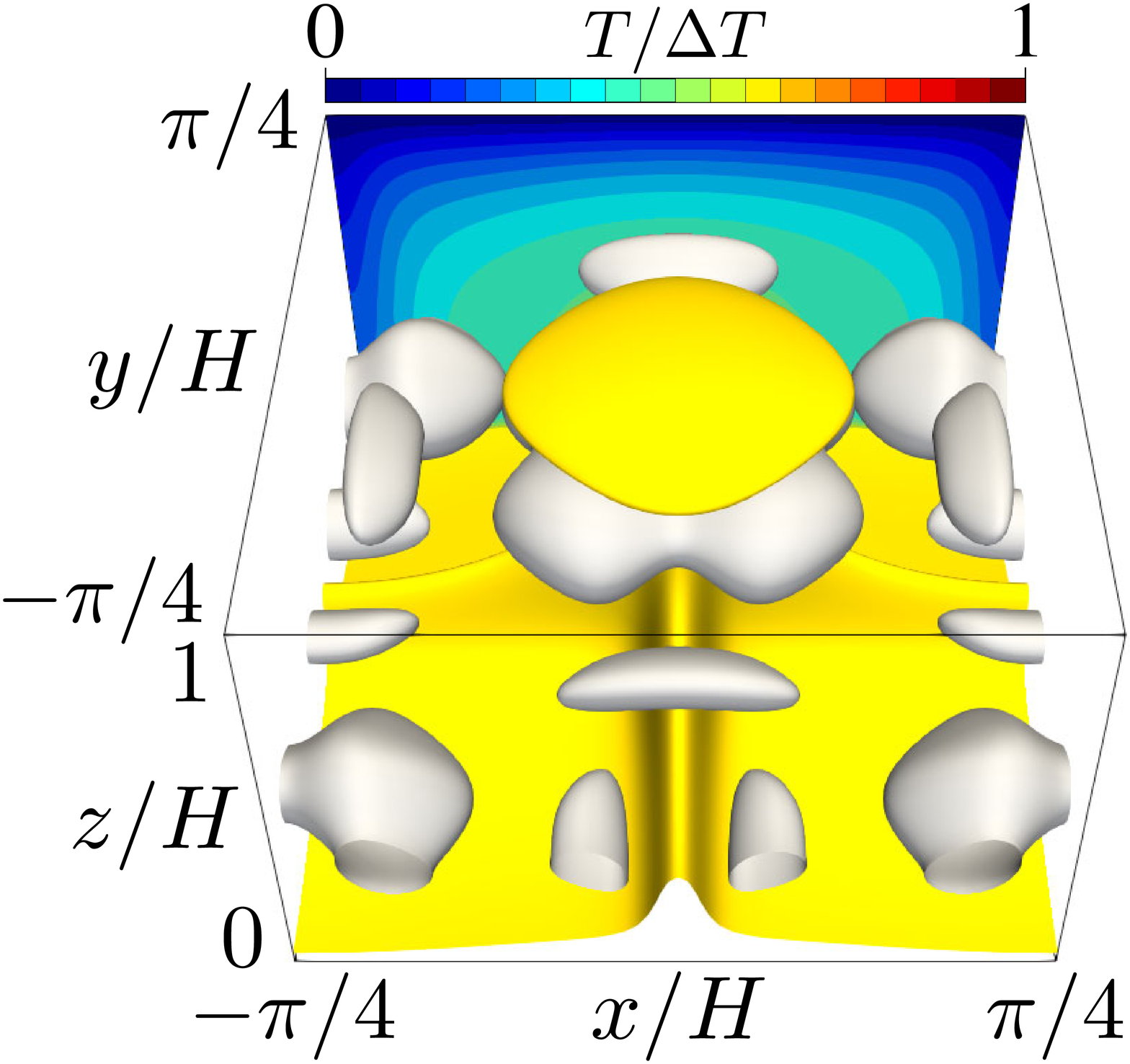}
	\end{minipage}
	\begin{minipage}{.32\linewidth}
	(\textit{b})\\
	\includegraphics[clip,width=\linewidth]{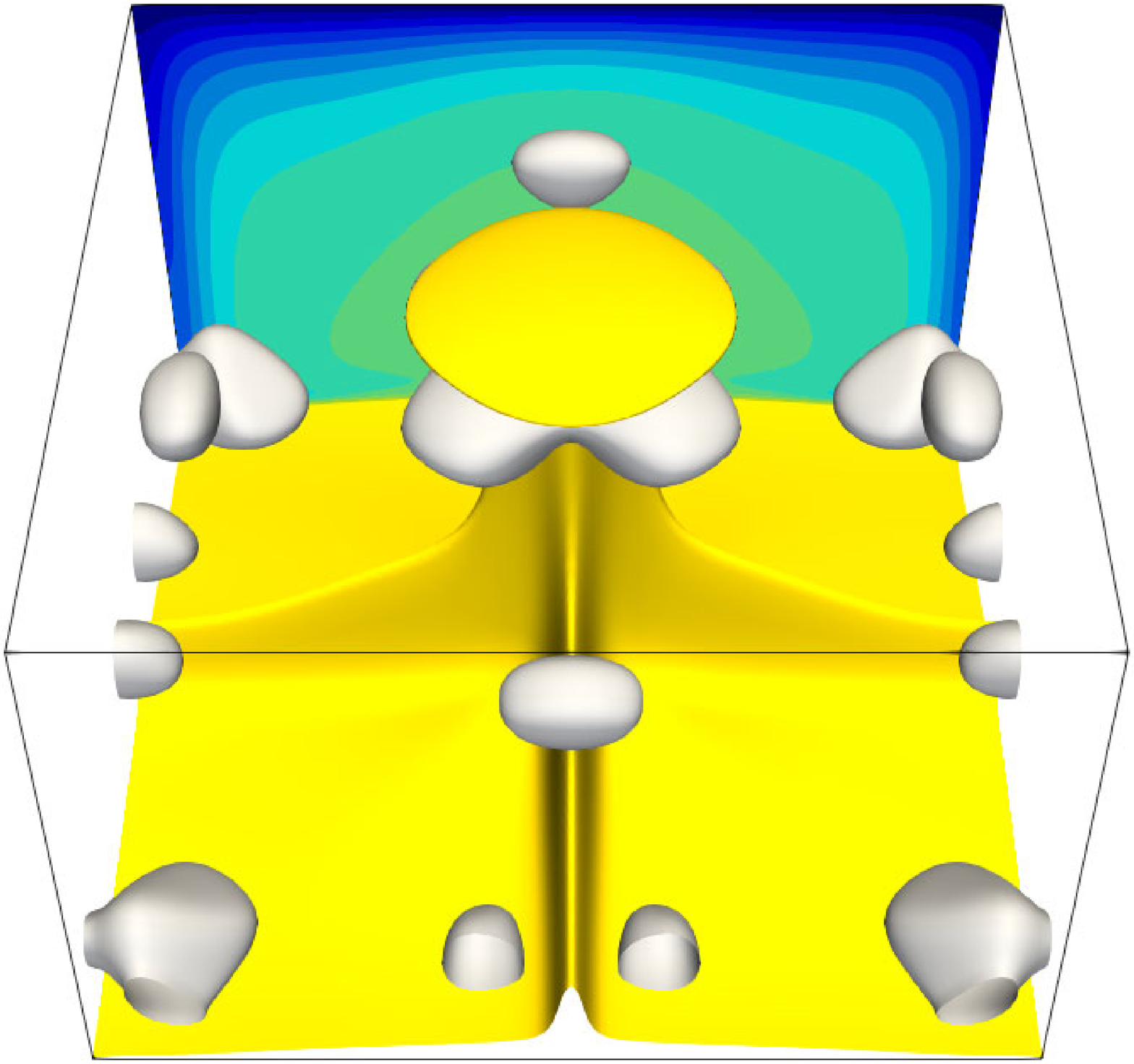}
        \end{minipage}
	\begin{minipage}{.32\linewidth}
	(\textit{c})\\
	\includegraphics[clip,width=\linewidth]{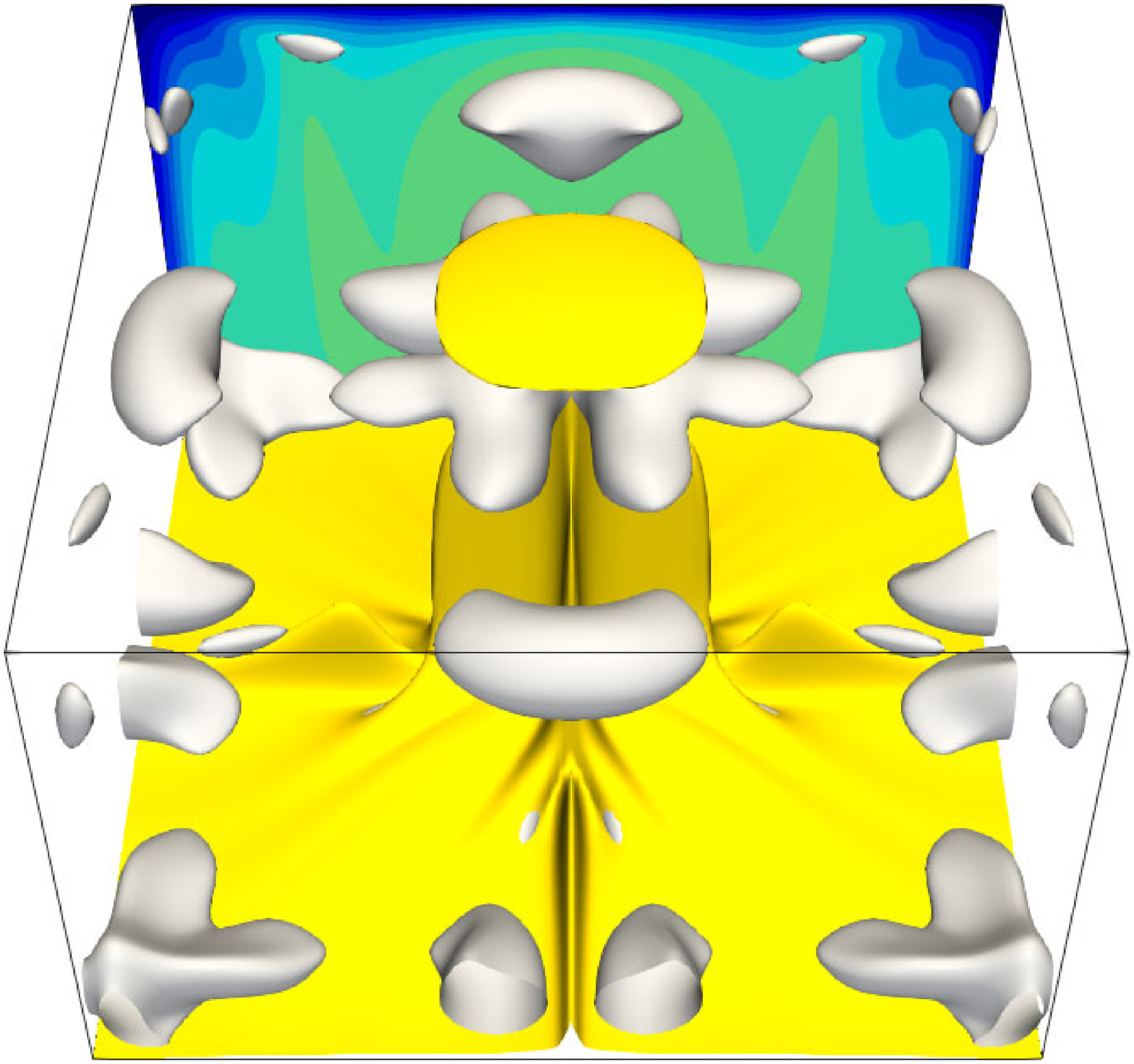}
	\end{minipage}
\caption{3D steady solution for $Pr=7$ in the domain with $L/H=\pi/2\approx1.57$ at (\textit{a}) $Ra=10^5$, (\textit{b}) $Ra=10^6$ and (\textit{c}) $Ra=10^7$.
The yellow and grey objects respectively show the isosurfaces of $T/\Delta T=0.6$ and (\textit{a}) $Q/(\kappa^2/H^4)=1\times10^5$, (\textit{b}) $Q/(\kappa^2/H^4)=2\times10^6$ and (\textit{c}) $Q/(\kappa^2/H^4)=2\times10^7$.
The contours represent $T$ in the plane $y/H=\pi/2$.
\label{fig:prandtl}}
\end{figure}

\section{Dependence of multi-scale steady solution on domain size, Prandtl number and spatial resolution}\label{sec:depend}
Figure \ref{fig:nu-ra_depend} shows the Nusselt number compensated by $Ra^{0.31}$ as a function of the Rayleigh number $Ra$ in the three-dimensional steady solutions for the different horizontal period $L$ and Prandtl number $Pr$.
The green, red and light blue symbols show the $L/H=2\pi/(k_{c}H)\approx2.02$, $\pi/2\approx1.57$ and $1$, respectively, for $Pr=1$, and the light red symbols represent $L/H=\pi/2$ for $Pr=7$, where $k_{c}=3.117/H$ is the wavenumber corresponding to the minimal critical $Ra_{c}=1708$ \citep{Drazin1981}.
Although we observe the dependence of $Nu$ on $Ra$, at $Ra\gtrsim10^{7}$ all the plots exhibit the values approximate to the optimal scaling $Nu-1=0.115Ra^{0.31}$ \citep[blue dashed,][]{Waleffe2015,Sondak2015} in the two-dimensional steady solutions.
We expect that the variation in the domain size and $Pr$ (for $1\lesssim Pr\lesssim10$) of $Nu$ would not be significant at high $Ra$, since the emergence of the small-scale plume and vortex structures near the walls, which are robustly observed in different $L$ and $Pr$ (figure \ref{fig:structure}, \ref{fig:aspect} and \ref{fig:prandtl}), might be key ingredient in the vertical heat flux.

In figure \ref{fig:nu-ra_depend} the symbols $+$, $\bullet$, $\times$ and $\circ$ show the results obtained on different grid points $(N_{x},N_{y},N_{z})=(64,64,64)$, $(128,128,128)$, $(192,192,128)$ and $(256,256,256)$, respectively, and the effects of the spatial resolutions on the $Nu$ are minor.
For our main results with $L/H=\pi/2$ and $Pr=1$, the grid points $(N_{x},N_{y},N_{z})=(128,128,128)$ are enough to evaluate the characteristics of the 3D steady solution at $Ra\lesssim10^{7}$; $(N_{x},N_{y},N_{z})=(256,256,256)$ are sufficient at $Ra\sim10^{7}$.
Furthermore, the Kolmogorov micro-scale length $\eta$ and the thermal conduction layer thickness $\delta$ in the 3D steady solution and the turbulent states at $Ra=10^{5},10^{6},10^{7}$ and $10^{7.42}\approx2.6\times10^{7}$ are shown in table \ref{table:details} together with the grid sizes.
Since the energy dissipation rate is a function of the wall-normal coordinate $z$, $\varepsilon(z)=(\nu/2){\langle {(\partial u_{i}/\partial x_{j}+\partial u_{j}/\partial x_{i})}^{2} \rangle}_{xyt}$, $\eta$ also depends on $z$.
${\langle \eta \rangle}_{z}$ and $\eta|_{c}$ are based on the total energy dissipation rate, ${\langle \varepsilon \rangle}_{z}$, and that at the centre of the fluid layer, $\varepsilon|_{z=H/2}$, respectively, and ${\langle \eta \rangle}_{z}<\eta|_{c}$ in all cases.
The grid size in the $x$-direction, $\Delta x(=\Delta y)$, and the maximal value of $z$, $\Delta z$, are comparable with $\eta$, and less than one third of $\delta$.
Therefore, the spatial resolution is sufficient to resolve the smallest-scale thermal and flow structures in the 3D steady solution and the turbulent states.
The present turbulent DNS data is obtained by averaging time of more than 200 convective time units based on the buoyancy-induced terminal velocity $U=\sqrt{g\alpha\Delta TH}$.

\begin{table}
  \begin{center}
  \def~{\hphantom{0}}
  \begin{tabular}{ccccccccc}
       & $Ra$ & $(N_{x},N_{y},N_{z})$ & $\Delta x/H$ & $\Delta z/H$ & ${\langle \eta \rangle}_{z}/H$ & $\eta|_{c}/H$ & $\delta/H$ & $\tau/(U/H)$\\
       & & & $\times10^{-2}$ & $\times10^{-2}$ & $\times10^{-2}$ & $\times10^{-2}$ & $\times10^{-2}$ & \\
\\
       & $10^{5}$  & $(128,128,128)$ & $1.22$ & $0.0153$--$1.24$ & $3.92$ & $4.25$ & $9.57$ & $-$\\
       3D steady& $10^{6}$  & $(128,128,128)$ & $1.22$ & $0.0153$--$1.24$ & $1.82$ & $2.08$ & $4.95$ & $-$\\
       solution& $10^{7}$  & $(256,256,256)$ & $0.611$ & $0.00379$--$0.616$ & $0.859$ & $1.01$ & $2.59$ & $-$\\
       & $10^{7.42}$  & $(256,256,256)$ & $0.611$ & $0.00379$--$0.616$ & $0.638$ & $0.746$ & $2.08$ & $-$\\
\\
       & $10^{5}$  & $(128,128,128)$ & $1.22$ & $0.0153$--$1.24$ & $3.99$ & $4.85$ & $10.1$ & $7906$\\
       Turbulent& $10^{6}$  & $(128,128,128)$ & $1.22$ & $0.0153$--$1.24$ & $1.88$ & $2.05$ & $5.57$ & $2500$\\
       states& $10^{7}$  & $(256,256,256)$ & $0.611$ & $0.00379$--$0.616$ & $0.897$ & $0.945$ & $3.04$ & $1028$\\
       & $10^{7.42}$  & $(256,256,256)$ & $0.611$ & $0.00379$--$0.616$ & $0.657$ & $0.685$ & $2.34$ & $513$\\
  \end{tabular}
  \caption{Numerical details of the 3D steady solution and the turbulent states for $Pr=1$ and $L/H=\pi/2$.
  $\Delta x$ and $\Delta z$ are the spatial resolutions in the $x$- and $z$-directions.
  ${\langle \eta \rangle}_{z}$ and $\eta|_{c}$ represent the Kolmogorov micro-scale length $\eta={(\nu^{3}/\varepsilon)}^{1/4}$ based on the vertical averaged energy dissipation rate, ${\langle \varepsilon \rangle}_{z}=(\nu/2){\langle {(\partial u_{i}/\partial x_{j}+\partial u_{j}/\partial x_{i})}^{2} \rangle}_{xyzt}$, and that at the centre of the fluid layer, $\varepsilon|_{z=H/2}$, respectively.
  $\delta$ is the thermal conduction layer thickness, $\delta/H=1/(2Nu)$.
  $\tau$ is the integral time to obtain the statistics, and $U=\sqrt{g\alpha\Delta T H}$ is the buoyancy-induced terminal velocity.
  \label{table:details}}
  \end{center}
\end{table}

\bibliography{ref}
\bibliographystyle{jfm}

\end{document}